\documentclass[iop]{emulateapj}

\newcommand{\sqcm}{${\rm cm^{-2}}$}
\newcommand{\cubcm}{${\rm cm^{-3}}$}
\newcommand{\mum}{${\rm \mu m}$}

\slugcomment{Accepted for publication in ApJ on 29 Aug 2013}

\shorttitle{Ice Formation and Grain Growth in Lupus}

\shortauthors{Boogert, et al.}

\received{May 15, 2013}

\begin{document}

\title{Infrared Spectroscopic Survey of the Quiescent Medium of Nearby
  Clouds: I. Ice Formation and Grain Growth in Lupus \altaffilmark{1}}

\author{A. C. A. Boogert\altaffilmark{2},
        J. E. Chiar\altaffilmark{3},
        C. Knez\altaffilmark{4,5},
        K. I. {\"O}berg\altaffilmark{6,7},
        L. G. Mundy\altaffilmark{4},
        Y. J. Pendleton\altaffilmark{8},
        A. G. G. M. Tielens\altaffilmark{9},
        E. F. van Dishoeck\altaffilmark{9, 10}}

\altaffiltext{1}{Based on observations made with ESO Telescopes at the La Silla Paranal Observatory under programme IDs 083.C-0942 and 085.C-0620.}
\altaffiltext{2}{IPAC, NASA Herschel Science Center,
  Mail Code 100-22, California Institute of Technology, Pasadena, CA
  91125, USA (email: aboogert@ipac.caltech.edu)}
\altaffiltext{3}{SETI Institute, Carl Sagan Center, 189 Bernardo Av.,
  Mountain View, CA 94043, USA}
\altaffiltext{4}{Department of Astronomy, University of Maryland,
  College Park, MD 20742, USA}
\altaffiltext{5}{Johns Hopkins University Applied Physics Laboratory,
  11100 Johns Hopkins Road, Laurel, MD 20723, USA}
\altaffiltext{6}{Departments of Chemistry and Astronomy, University of
  Virginia, Charlottesville, VA 22904, USA}
\altaffiltext{7}{Harvard-Smithsonian Center for Astrophysics, 60
  Garden Street, Cambridge, MA 02138, USA}
\altaffiltext{8}{Solar System Exploration Research Virtual Institute,
  NASA Ames Research Center, Moffett Field, CA 94035, USA}
\altaffiltext{9}{Leiden Observatory, Leiden University, PO Box 9513,
  2300 RA Leiden, the Netherlands}
\altaffiltext{10}{Max Planck Institut f\"ur Extraterrestrische Physik
  (MPE), Giessenbachstr. 1, 85748 Garching, Germany}

\begin{abstract}

Infrared photometry and spectroscopy (1-25 \mum) of background stars
reddened by the Lupus molecular cloud complex are used to determine
the properties of the grains and the composition of the ices before
they are incorporated into circumstellar envelopes and disks. H$_2$O
ices form at extinctions of $A_{\rm K}=0.25\pm 0.07$ mag ($A_{\rm
  V}=2.1\pm 0.6$). Such a low ice formation threshold is consistent
with the absence of nearby hot stars.  Overall, the Lupus clouds are
in an early chemical phase.  The abundance of H$_2$O ice (2.3$\pm 0.1
\times 10^{-5}$ relative to $N_{\rm H}$) is typical for quiescent
regions, but lower by a factor of 3-4 compared to dense envelopes of
YSOs.  The low solid CH$_3$OH abundance ($<3-8$\% relative to H$_2$O)
indicates a low gas phase H/CO ratio, which is consistent with the
observed incomplete CO freeze out. Furthermore it is found that the
grains in Lupus experienced growth by coagulation. The mid-infrared
($>5$ \mum) continuum extinction relative to $A_{\rm K}$ increases as
a function of $A_{\rm K}$. Most Lupus lines of sight are well fitted
with empirically derived extinction curves corresponding to $R_{\rm
  V}\sim 3.5$ ($A_{\rm K}=0.71$) and $R_{\rm V}\sim 5.0$ ($A_{\rm
  K}=1.47$).  For lines of sight with $A_{\rm K}>1.0$ mag, the
$\tau_{9.7}/A_{\rm K}$ ratio is a factor of 2 lower compared to the
diffuse medium. Below 1.0 mag, values scatter between the dense and
diffuse medium ratios. The absence of a gradual transition between
diffuse and dense medium-type dust indicates that local conditions
matter in the process that sets the $\tau_{9.7}/A_{\rm K}$ ratio. This
process is likely related to grain growth by coagulation, as traced by
the $A_{7.4}/A_{\rm K}$ continuum extinction ratio, but not to ice
mantle formation.  Conversely, grains acquire ice mantles before the
process of coagulation starts.

\end{abstract}


\section{Introduction}~\label{sec:intro}

Dense cores and clouds are the birthplaces of stars and their
planetary systems (e.g., \citealt{eva09}), and it is therefore
important to know their composition.  Gas phase abundances are
strongly reduced in these environments as species freeze out onto
grains (CO, CS; \citealt{ber01}), and new molecules are formed by
grain surface chemistry (e.g., H$_2$O, CH$_4$, CO$_2$,
\citealt{tie82}). Infrared spectroscopy of the vibrational absorption
bands of ices against the continuum emission of background stars is
thus a powerful tool to determine the composition of dense media
\citep{whi83}.

The Taurus Molecular Cloud (TMC) is the first cloud in which frozen
H$_2$O \citep{whi83}, CO \citep{whi85}, and CO$_2$ \citep{whi98} were
detected using background stars.  This is also the case for the 6.85
\mum\ band \citep{kne05}, whose carrier is uncertain (possibly
NH$_4^+$).  Recently solid CH$_3$OH was discovered toward several
isolated dense cores \citep{boo11,chi11}.  These and follow up studies
showed that the ice abundances depend strongly on the environment. The
extinction threshold for H$_2$O ice formation is a factor of two
higher for the Ophiuchus (Oph) cloud than it is for TMC ($A_{\rm
  V}=$10-15 versus 3.2 mag; \citealt{tan90, whi01}).  These variations
may reflect higher local interstellar radiation fields (e.g., hot
stars in the Oph neighborhood), which remove H$_2$O and its precursors
from the grains either by photodesorption or sublimation at the cloud
edge \citep{hol09}. Deeper in the cloud, ice mantle formation may be
suppressed by shocks and radiation fields from Young Stellar Objects
(YSOs), depending on the star formation rate (SFR) and initial mass
function (IMF).  The latter may apply in particular to the freeze out
of the volatile CO species, and, indirectly, to CH$_3$OH. CO freeze
out sets the gas phase H/CO ratio, which sets the CH$_3$OH formation
rate \citep{cup09}. High CH$_3$OH abundances may be produced on time
scales that depend on dust temperature and other local conditions
(e.g., shocks). This may well explain the observed large CH$_3$OH
abundance variations: $N$(CH$_3$OH)/$N$(H$_2$O)$\leq 3$\% toward TMC
background stars \citep{chi95} and $\sim 10$\% toward some isolated
dense cores \citep{boo11}.

The number of dense clouds and the number of sight-lines within each
cloud observed with mid-infrared spectroscopy ($\lambda>5$ \mum) is
small. Ice and silicate inventories were determined toward four TMC
and one Serpens cloud background stars \citep{kne05}. Many more lines
of sight were recently studied toward isolated dense cores
\citep{boo11}.  These cores have different physical histories and
conditions, however, and may not be representative of dense clouds.
Their interstellar media lack the environmental influences of outflows
and the resulting turbulence often thought to dominate in regions of
clustered star formation (clouds).  Their star formation time scales
can be significantly larger due to the dominance of magnetic fields
over turbulence and the resulting slow process of ambipolar diffusion
\citep{shu87, eva99}. Over time, the ice composition likely reflects
the physical history of the environment.  This, in turn may be
preserved in the ices in envelopes, disks and planetary
systems. Quiescent cloud and core ices are converted to more complex
organics in YSO envelopes \citep{obe11}.  2D collapse models of YSOs
show that subsequently all but the most volatile envelope ices (CO,
N$_2$) survive the infall phase up to radii $>10$ AU from the star
\citep{vis09}.  They are processed at radii $<30$ AU, further
increasing the chemical complexity.  For this reason, it is required
to determine the ice abundances in a larger diversity of quiescent
environments.  A Spitzer spectroscopy program (P.I. C. Knez) was
initiated to observe large samples of background stars, selected from
2MASS and Spitzer photometric surveys of the nearby Lupus, Serpens,
and Perseus Molecular Clouds. This paper focuses on the Lupus
cloud. Upcoming papers will present mid-infrared spectroscopy of stars
behind the Serpens and Perseus clouds.

The Lupus cloud complex is one of the main nearby low mass star
forming regions. It is located near the Scorpius-Centaurus OB
association and consists of a loosely connected group of clouds
extended over $\sim$20 degrees (e.g., \citealt{com08}). The Lupus I,
III, and IV clouds were mapped with Spitzer/IRAC and MIPS broad band
filters, and analyzed together with 2MASS near infrared maps
\citep{eva09}. In this paper, only stars behind the Lupus I and IV
clouds will be studied. This is the first study of Lupus background
stars. Compared to other nearby clouds, the Lupus clouds likely
experienced less impact from nearby massive stars and internal YSOs.
While OB stars in the Scorpius-Centaurus association may have
influenced the formation of the clouds, they are relatively far away
($\sim 17$ pc) and their current impact on the Lupus clouds is most
likely smaller compared to that of massive stars on the Oph, Serpens,
and Perseus clouds \citep{eva09}.  Likewise, star formation within the
Lupus clouds is characterized by a relatively low SFR, 0.83
M$_\odot$/Myr/pc$^2$, versus 1.3, 2.3, and 3.2 for Perseus, Ophiuchus
and the Serpens clouds, respectively \citep{eva09}. In addition, the
mean stellar mass of the YSOs (0.2 M$_\odot$; \citealt{mer08}) is low
compared to that of other clouds (e.g., Serpens 0.7 M$_\odot$) as well
as to that of the IMF (0.5 M$_\odot$).  Lupus also stands out with a
low fraction of Class I YSOs \citep{eva09}. Within Lupus, the
different clouds have distinct characteristics.  While Herschel
detections of prestellar cores and Class 0 YSOs indicate that both
Lupus I and IV have an increasing SFR, star formation in Lupus IV has
just begun, considering its low number of prestellar sources
\citep{ryg12}.  The Lupus IV cloud is remarkable in that the
Spitzer-detected YSOs are distributed away from the highest extinction
regions.  Extinction maps produced by the c2d team show that Lupus IV
contains a distinct extinction peak, while Lupus I has a lower, more
patchy extinction ($A_{\rm V}$=32.6 versus 26.5 mag at a resolution of
120$''$).  It is comparable to the Serpens cloud (33.5 mag at 120$''$
resolution), but factors of 1.5-2 lower compared to the Perseus and
Ophiuchus clouds.

Both volatiles and refractory dust can be traced in the mid-infrared
spectra of background stars.  This paper combines the study of ice and
silicate features with line of sight extinctions. The ice formation
threshold toward Lupus is investigated as well as the relation of the
9.7 \mum\ band of silicates with the continuum extinction.  The 9.7
\mum\ band was extensively studied toward background stars tracing
dense clouds and cores.  While no differences were found between
clouds and cores, its strength and shape are distinctly different
compared to the diffuse ISM.  The peak optical depth of the
9.7~\mum\ band relative to the K-band continuum extinction is a factor
of $\sim$2 smaller in dense lines of sight
\citep{chi07,boo11,chi11}. The short wavelength wing is also more
pronounced. Grain growth cannot explain these effects simultaneously
\citep{bre11}. On the other hand, the same spectra of background stars
show grain growth by increased continuum extinction at longer
wavelengths (up to at least 25~\mum; \citealt{mcc09, boo11}), in
agreement with broad band studies \citep{cha09}.

The selection of the background stars is described in \S\ref{sec:sou},
and the reduction of the ground-based and {\it Spitzer} spectra in
\S\ref{sec:obs}. In \S\ref{sec:cont}, the procedure to fit the stellar
continua is presented, a crucial step in which ice and silicate
features are separated from stellar features and continuum
extinction. Subsequently, in \S\ref{sec:col}, the peak and integrated
optical depths of the ice and dust features are derived, as well as
column densities for the identified species. Then in \S\ref{sec:60},
the derived parameters $A_{\rm K}$, $\tau_{3.0}$, and $\tau_{9.7}$ are
correlated with each other.  \S\ref{sec:thresh} discusses the Lupus
ice formation threshold and how it compares to other clouds.  The
slope of the $A_{\rm K}$ versus $\tau_{3.0}$ relations is discussed in
\S\ref{sec:slope}.  The ice abundances are put into context in
\S\ref{sec:complex}. The $A_{\rm K}$ versus $\tau_{9.7}$ relation, and
in particular the transition from diffuse to dense cloud values, is
discussed in \S\ref{sec:tauak}. Finally, the conclusions are
summarized and an outlook to future studies is presented in
\S\ref{sec:concl}.

\begin{deluxetable}{llrll}
\tabletypesize{\scriptsize}
\tablecolumns{5}
\tablewidth{0pc}
\tablecaption{Source Sample~\label{t:sample}}
\tablehead{
\colhead{Source}& \colhead{Cloud} & \colhead{AOR key\tablenotemark{a}} & \colhead{Module\tablenotemark{b}} & \colhead{$\lambda_{\rm NIR}$\tablenotemark{c}}\\
\colhead{2MASS~J}& \colhead{   }   & \colhead{ } & \colhead{ } & \colhead{\mum}\\}
\startdata
15382645$-$3436248                  & Lup I            &23077120 &SL, LL2  & 1.88-4.17\\
15423699$-$3407362                  & Lup I            &23078400 &SL, LL2  & 1.88-4.17\\
15424030$-$3413428${\rm ^d}$        & Lup I            &23077888 &SL, LL2  & 1.88-4.17\\
15425292$-$3413521                  & Lup I            &23077632 &SL, LL2  & 1.88-5.06\\
15444127$-$3409596                  & Lup I            &23077376 &SL, LL   & 1.88-4.17\\
15450300$-$3413097                  & Lup I            &23077376 &SL, LL   & 1.88-5.06\\
15452747$-$3425184                  & Lup I            &23077888 &SL, LL2  & 1.88-5.06\\
15595783$-$4152396                  & Lup IV           &23079168 &SL, LL2  & 1.88-4.17\\
16000067$-$4204101                  & Lup IV           &23079680 &SL, LL2  & 1.88-4.17\\
16000874$-$4207089                  & Lup IV           &23078912 &SL, LL   & 1.88-4.17\\
16003535$-$4209337                  & Lup IV           &23081216 &SL, LL2  & 1.88-4.17\\
16004226$-$4146411                  & Lup IV           &23079424 &SL, LL2  & 1.88-4.17\\
16004739$-$4203573                  & Lup IV           &23082240 &SL, LL2  & 1.88-4.17\\
16004925$-$4150320                  & Lup IV           &23079936 &SL, LL   & 1.88-5.06\\
16005422$-$4148228                  & Lup IV           &23079936 &SL, LL   & 1.88-4.17\\
16005511$-$4132396                  & Lup IV           &23078656 &SL, LL   & 1.88-4.17\\
16005559$-$4159592                  & Lup IV           &23079680 &SL, LL2  & 1.88-4.17\\
16010642$-$4202023                  & Lup IV           &23081984 &SL, LL2  & 1.88-4.17\\
16011478$-$4210272                  & Lup IV           &23079168 &SL, LL2  & 1.88-4.17\\
16012635$-$4150422                  & Lup IV           &23081728 &SL, LL2  & 1.88-4.17\\
16012825$-$4153521                  & Lup IV           &23081472 &SL, LL2  & 1.88-4.17\\
16013856$-$4133438                  & Lup IV           &23079424 &SL, LL2  & 1.88-4.17\\
16014254$-$4153064                  & Lup IV           &23082496 &SL, LL2  & 1.88-5.06\\
16014426$-$4159364                  & Lup IV           &23080192 &SL, LL2  & 1.88-4.17\\
16015887$-$4141159                  & Lup IV           &23078656 &SL, LL   & 1.88-4.17\\
16021102$-$4158468                  & Lup IV           &23080192 &SL, LL2  & 1.88-5.06\\
16021578$-$4203470                  & Lup IV           &23078656 &SL, LL   & 1.88-4.17\\
16022128$-$4158478                  & Lup IV           &23080704 &SL, LL2  & 1.88-4.17\\
16022921$-$4146032                  & Lup IV           &23078912 &SL, LL   & 1.88-4.17\\
16023370$-$4139027                  & Lup IV           &23080960 &SL, LL2  & 1.88-4.17\\
16023789$-$4138392                  & Lup IV           &23080448 &SL, LL2  & 1.88-4.17\\
16024089$-$4203295                  & Lup IV           &23080448 &SL, LL2  & 1.88-4.17\\
\enddata
\tablenotetext{a}{Identification number for {\it Spitzer} observations}
\tablenotetext{b}{{\it Spitzer}/IRS modules used: SL=Short-Low (5-14~\mum, $R\sim100$), LL2=Long-Low 2 (14-21.3~\mum, $R\sim100$), LL=Long-Low 1 and 2 (14-35~\mum, $R\sim100$)}
\tablenotetext{c}{Wavelength coverage of complementary near-infrared ground-based observations, excluding the ranges $\sim$2.55-2.85, and $\sim$4.15-4.49~\mum\ blocked by the Earth's atmosphere.}
\tablenotetext{d}{This is not a background star, but rather a Class III YSO within Lupus I \citep{mer08}. The ice and dust features will be derived in this work, but they will be omitted from subsequent analysis.}
\end{deluxetable}

\section{Source Selection}~\label{sec:sou}

Background stars were selected from the Lupus I and IV clouds that
were mapped with {\it Spitzer}/IRAC and MIPS by the c2d Legacy team
\citep{eva03, eva07}. The maps are complete down to $A_{\rm V}$=3 and
$A_{\rm V}$=2 for Lupus I and IV, respectively \citep{eva03}.  The
selected sources have an overall SED (2MASS 1-2~\mum, IRAC 3-8~\mum,
MIPS 24~\mum) of a reddened Rayleigh-Jeans curve. They fall in the
``star'' category in the c2d catalogs and have MIPS 24 \mum\ to IRAC 8
\mum\ flux ratios greater than 4.  In addition, fluxes are high enough
($>$10~mJy at 8.0~\mum) to obtain {\it Spitzer}/IRS spectra of high
quality (S/N$>$50) within $\sim$20 minutes of observing time per
module. This is needed to detect the often weak ice absorption
features and determine their shapes and peak positions.  This resulted
in roughly 100 stars behind Lupus I and IV. The list was reduced by
selecting $\sim$10 sources in each interval of $A_{\rm V}$: 2-5, 5-10,
and $>$10 mag (taking $A_{\rm V}$ from the c2d catalogs) and making
sure that the physical extent of the cloud is covered.  The final list
contains nearly all high $A_{\rm V}$ lines of sight. At low
extinctions, many more sources were available and the brightest were
selected.  The observed sample of 25 targets toward Lupus IV, and 7
toward Lupus I is listed in Table~\ref{t:sample}. The analysis showed
that the SEDs of three Lupus I and two Lupus IV sources cannot be
fitted with stellar models (\S\ref{sec:cont}).  One of these is a
confirmed Class III ``cold disk'' YSO (2MASS J$15424030-3413428$;
\citealt{mer08}).  The ice and dust feature strengths and abundances
are derived for these five sources, but they are not used in the
quiescent medium analysis.

\begin{figure*}
\includegraphics[angle=90, scale=0.51]{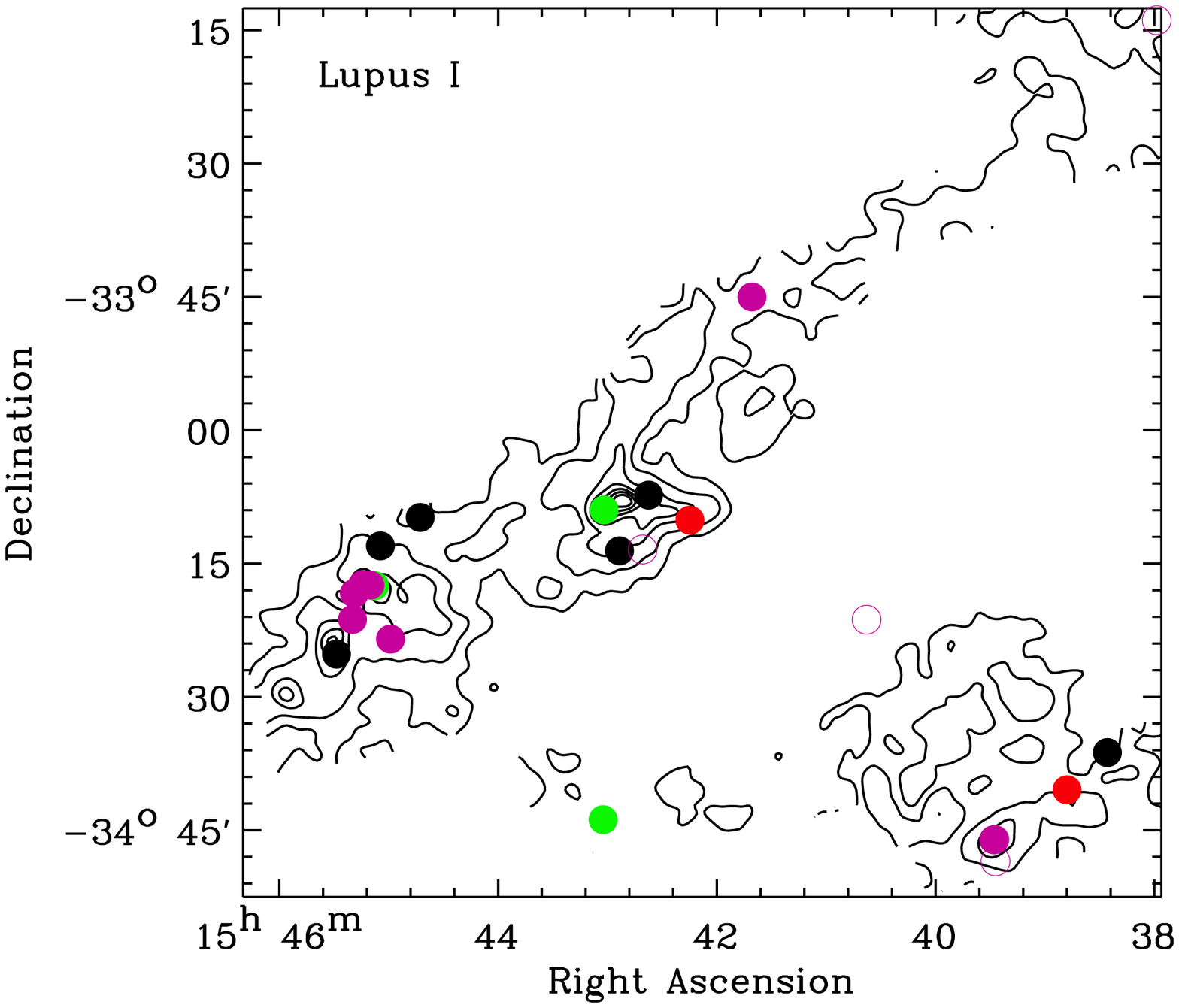}
\includegraphics[angle=90, scale=0.51]{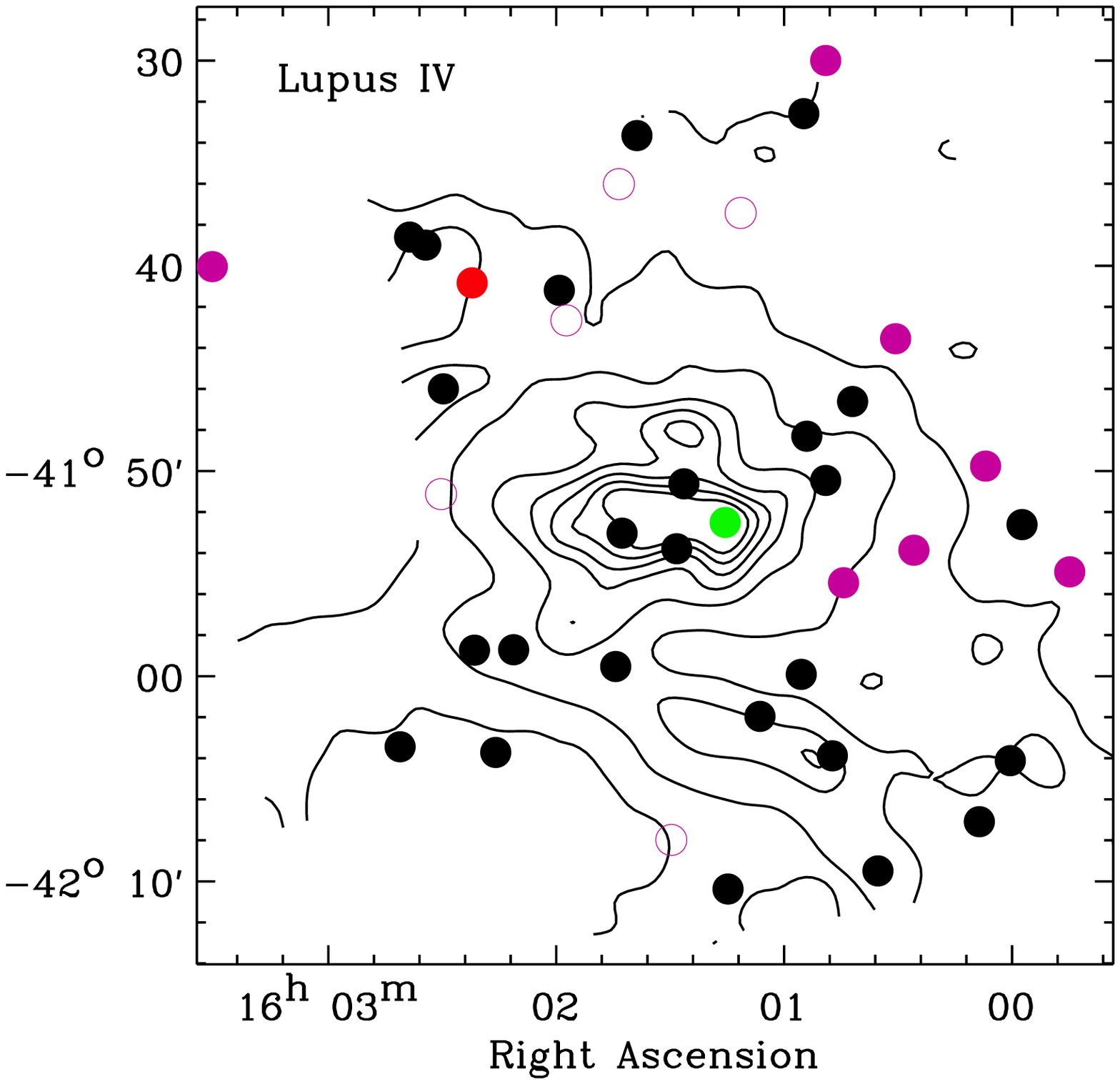}
\caption{Background stars (black filled circles) observed toward the
  Lupus I (left panel) and Lupus IV (right) clouds, over-plotted on
  extinction maps \citep{eva07} with contours
  $A_K$=0.25,0.5,1.0,1.5,2.0,2.5,3.0 mag (assuming $A_{\rm V}/A_{\rm
    K}=7.6$ for $R_{\rm V}=5.0$; \citealt{car89}). The red filled
  circles indicate Class I YSOs, green filled circles Flat spectrum
  YSOs, purple filled circles Class II YSOs, and purple open circles
  Class III YSOs (all from \citealt{mer08}).}~\label{f:maps}
\end{figure*}

Figure~\ref{f:maps} plots the location of the observed background
stars on extinction maps derived from 2MASS and Spitzer photometry
\citep{eva07}. The maps also show all YSOs identified in the Spitzer
study of \citet{mer08}. Some lines of sight are in the same area as
Class I and Flat spectrum sources, but not closer than a few
arcminutes.

\section{Observations and Data Reduction}~\label{sec:obs}

{\it Spitzer}/IRS spectra of background stars toward the Lupus I and
IV clouds were obtained as part of a dedicated Open Time program (PID
40580). Table~\ref{t:sample} lists all sources with their AOR keys,
and the IRS modules they were observed in. The SL module, covering the
5-14 \mum\ range, includes several ice absorption bands as well as the
9.7 \mum\ band of silicates, and had to highest signal-to-noise goal
($>$50). The LL2 module (14-21 \mum) was included to trace the 15
\mum\ band of solid CO$_2$ and for a better overall continuum
determination, although at a lower signal-to-noise ratio of $>30$. At
longer wavelengths, the background stars are weaker, and the LL1
module ($\sim$20-35 \mum) was used for only $\sim 30$\% of the
sources.  The spectra were extracted and calibrated from the
two-dimensional Basic Calibrated Data produced by the standard {\it
  Spitzer} pipeline (version S16.1.0), using the same method and
routines discussed in \citet{boo11}. Uncertainties (1$\sigma$) for
each spectral point were calculated using the ``func'' frames provided
by the {\it Spitzer} pipeline.

The {\it Spitzer} spectra were complemented by ground-based VLT/ISAAC
\citep{moo98} K and L-band spectra. Six bright sources were also
observed in the M-band. The observations were done in ESO programs
083.C-0942(A) (visitor mode) and 085.C-0620(A) (service mode) spread
over the time frame of 25 June 2009 until 14 August 2010.  The K-band
spectra were observed in the SWS1-LR mode with a slit width of 0.3'',
yielding a resolving power of $R=$1500. Most L- and M-band spectra
were observed in the LWS3-LR mode with a slit width of 0.6'', yielding
resolving powers of $R=$600 and 800, respectively.  The ISAAC pipeline
products from the ESO archive could not be used for scientific
analysis because of errors in the wavelength scale (the lamp lines
were observed many hours from the sky targets). Instead, the data were
reduced from the raw frames in a way standard for ground-based
long-slit spectra with the same IDL routines used for Keck/NIRSPEC
data previously \citep{boo08}.  Sky emission lines were used for the
wavelength calibration and bright, nearby main sequence stars were
used as telluric and photometric standards.  The final spectra have
higher signal-to-noise ratios than the final ESO/ISAAC pipeline
spectra because the wavelength scale of the telluric standards were
matched to the science targets before division, using sky emission
lines as a reference.

In the end, all spectra were multiplied along the flux scale in order
to match broad-band photometry from the 2MASS \citep{skr06}, {\it
  Spitzer} c2d \citep{eva07}, and WISE \citep{wri10} surveys using the
appropriate filter profiles.  The same photometry is used in the
continuum determination discussed in \S\ref{sec:cont}.  Catalog flags
were taken into account, such that the photometry of sources listed as
being confused within a 2\arcsec\ radius or being located within
2\arcsec\ of a mosaic edge were treated as upper limits.  The c2d
catalogs do not include flags for saturation. Therefore, photometry
exceeding the IRAC saturation limits (at the appropriate integration
times) were flagged as lower limits. In those cases, the nearby WISE
photometric points were used instead.  Finally, as the relative
photometric calibration is important for this work, the uncertainties
in the {\it Spitzer} c2d and 2MASS photometry were increased with the
zero-point magnitude uncertainties listed in Table 21 of \citet{eva07}
and further discussed in their \S3.5.3.


\section{Results}~\label{sec:res}

The observed spectra (left panels of Fig.~\ref{f:obs1}) show the
distinct 3.0 and 9.7~\mum\ absorption features of H$_2$O ice and
silicates on top of reddened stellar continua.  These are the first
detections of ices and silicates in the quiescent medium of the Lupus
clouds.  The weaker 6.0, 6.85, and 15 \mum\ ice bands are evident
after a global continuum is subtracted from the spectra (right panels
of Fig.~\ref{f:obs1}).  Features from the stellar photosphere are
present as well (e.g., 2.4 and 8.0~\mum).  The separation of
interstellar and photospheric features is essential for this work and
is discussed next.

\begin{figure*}
\includegraphics[angle=90, scale=0.75]{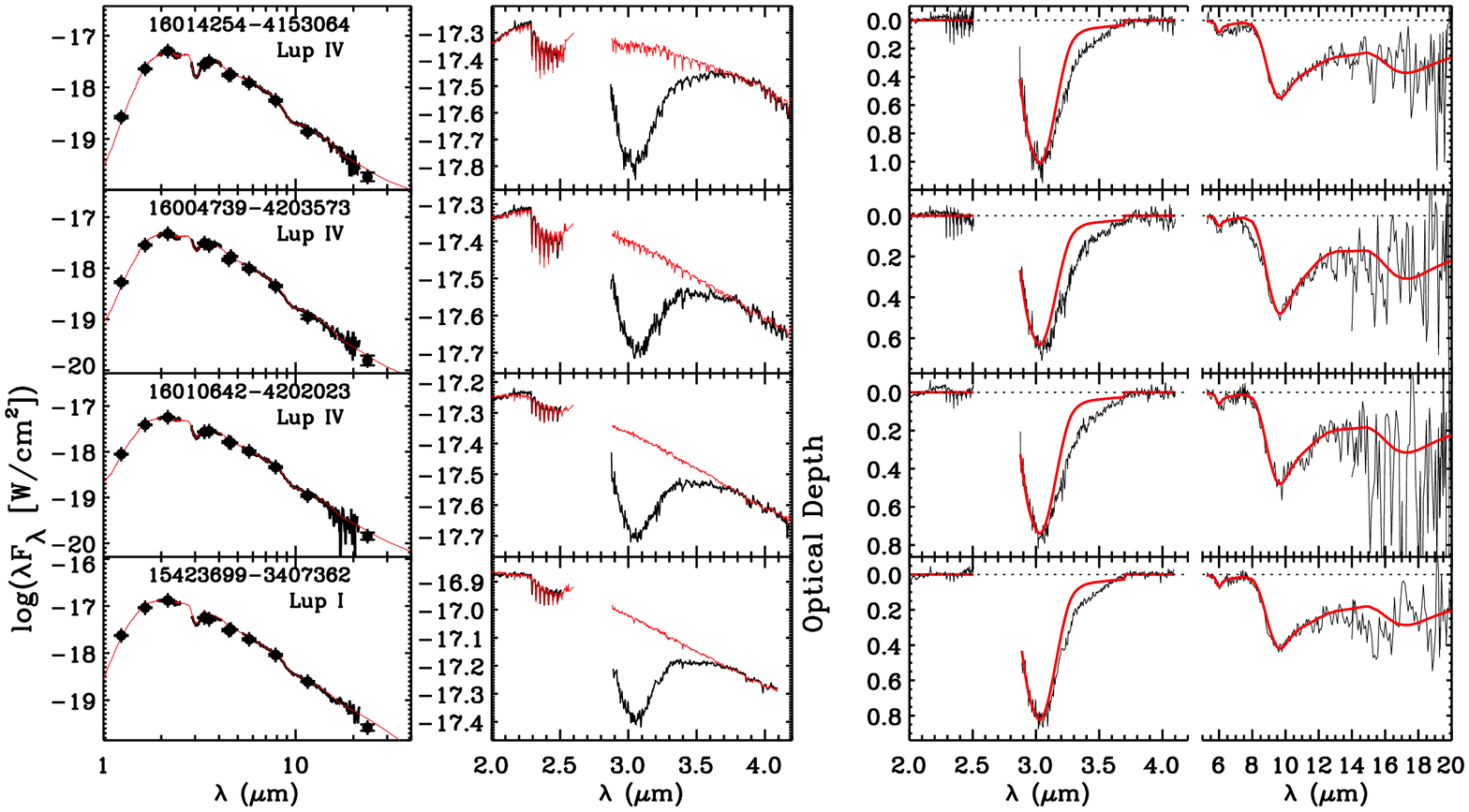}
\includegraphics[angle=90, scale=0.75]{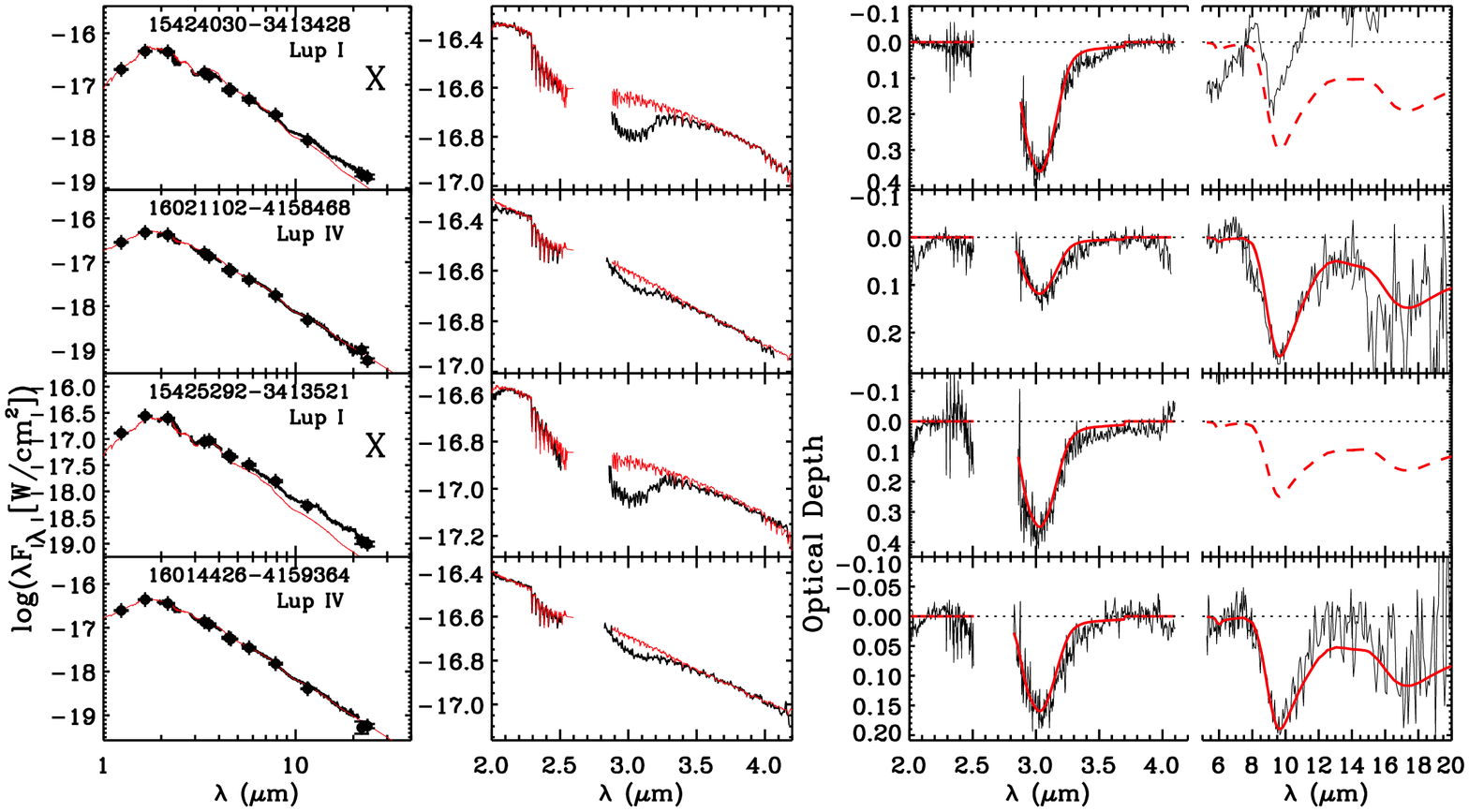}
\caption{{\bf Left panels:} Observed ground-based and {\it
    Spitzer}/IRS spectra combined with broad band photometry (filled
  circles), and lower limits (open triangles) and 3$\sigma$ upper
  limits (open circles) thereof. The red lines represent the fitted
  models, using synthetic stellar spectra (\S\ref{sec:cont}). The
  sources are sorted in decreasing ${\rm A_K}$ values from top to
  bottom. Sources labeled with 'X' have poor long wavelength fits and
  will not be further treated as background stars. {\bf Middle
    panels:} Observed ground-based $K$ and $L$-band spectra. The red
  lines represent the fitted models (excluding H$_2$O ice), using
  stellar spectra from the IRTF database \citep{ray09}. {\bf Right
    panels:} Optical depth spectra, derived using the continuum models
  of the left panels. The red lines indicate the modeled H$_2$O ice
  and silicates spectra. For clarity, error bars of the spectral data
  points are not shown.}~\label{f:obs1}
\end{figure*}

\setcounter{figure}{1}

\begin{figure*}
\includegraphics[angle=90, scale=0.75]{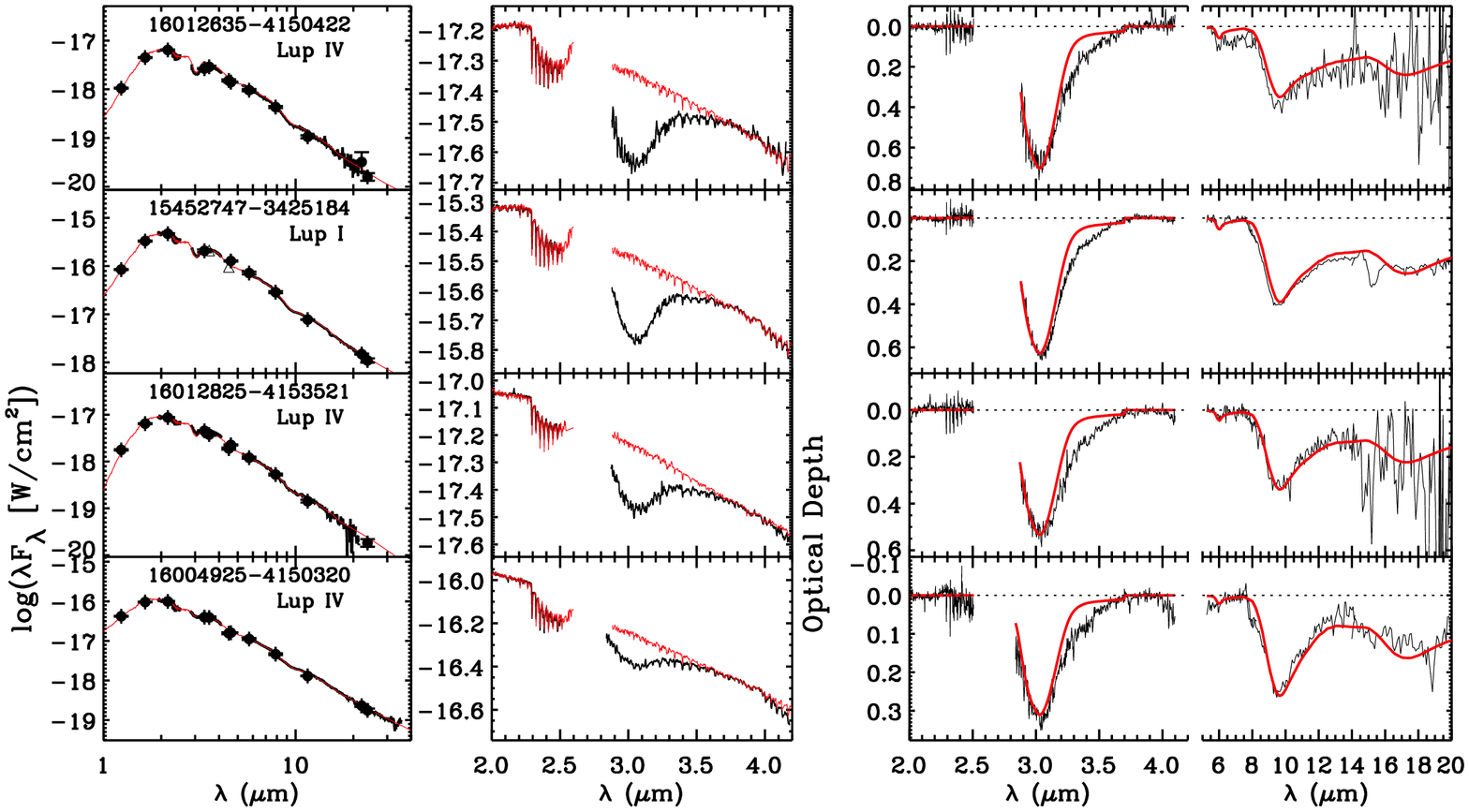}
\includegraphics[angle=90, scale=0.75]{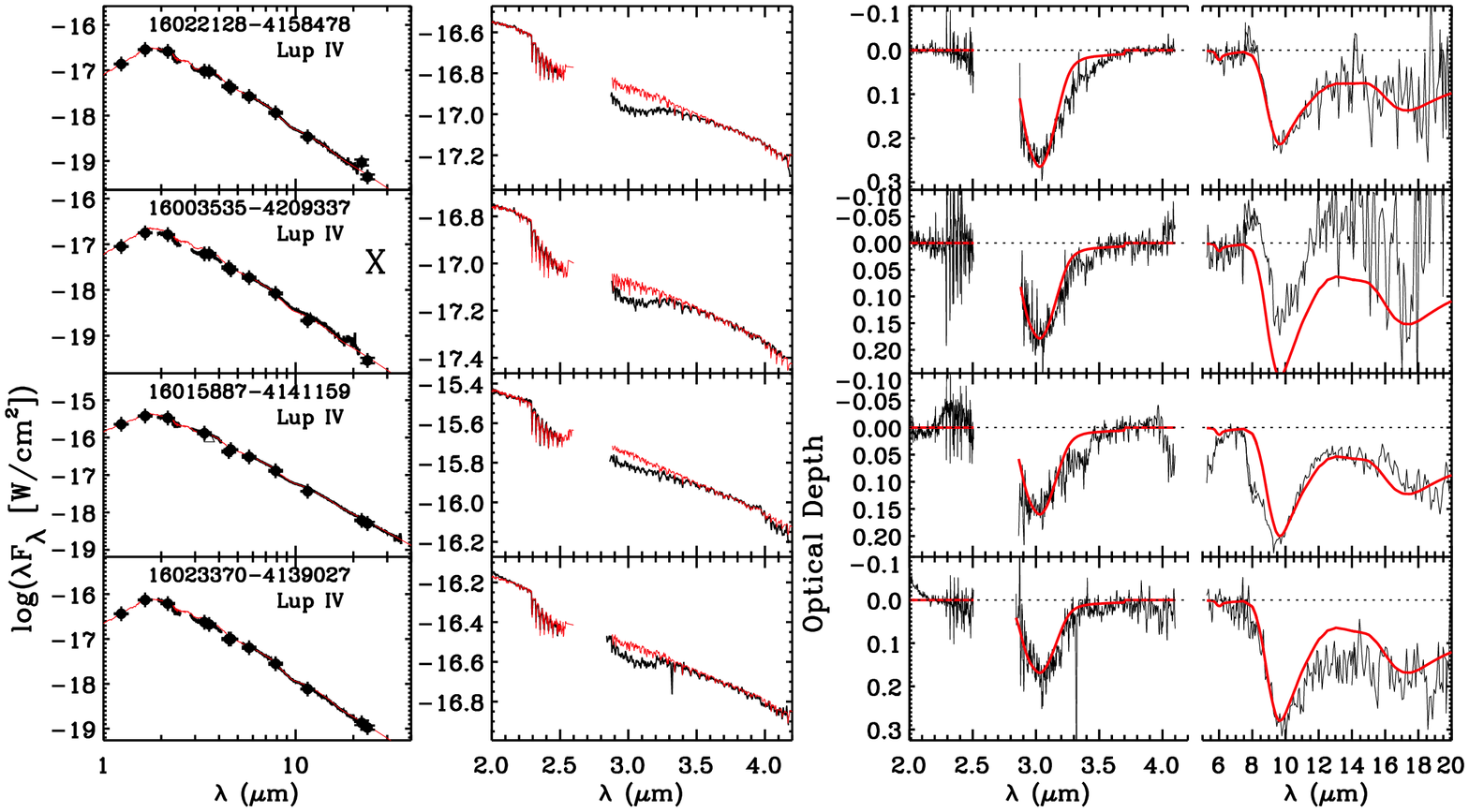}
\caption{(Continuation)}~\label{f:obs3}
\end{figure*}

\setcounter{figure}{1}

\begin{figure*}
\includegraphics[angle=90, scale=0.75]{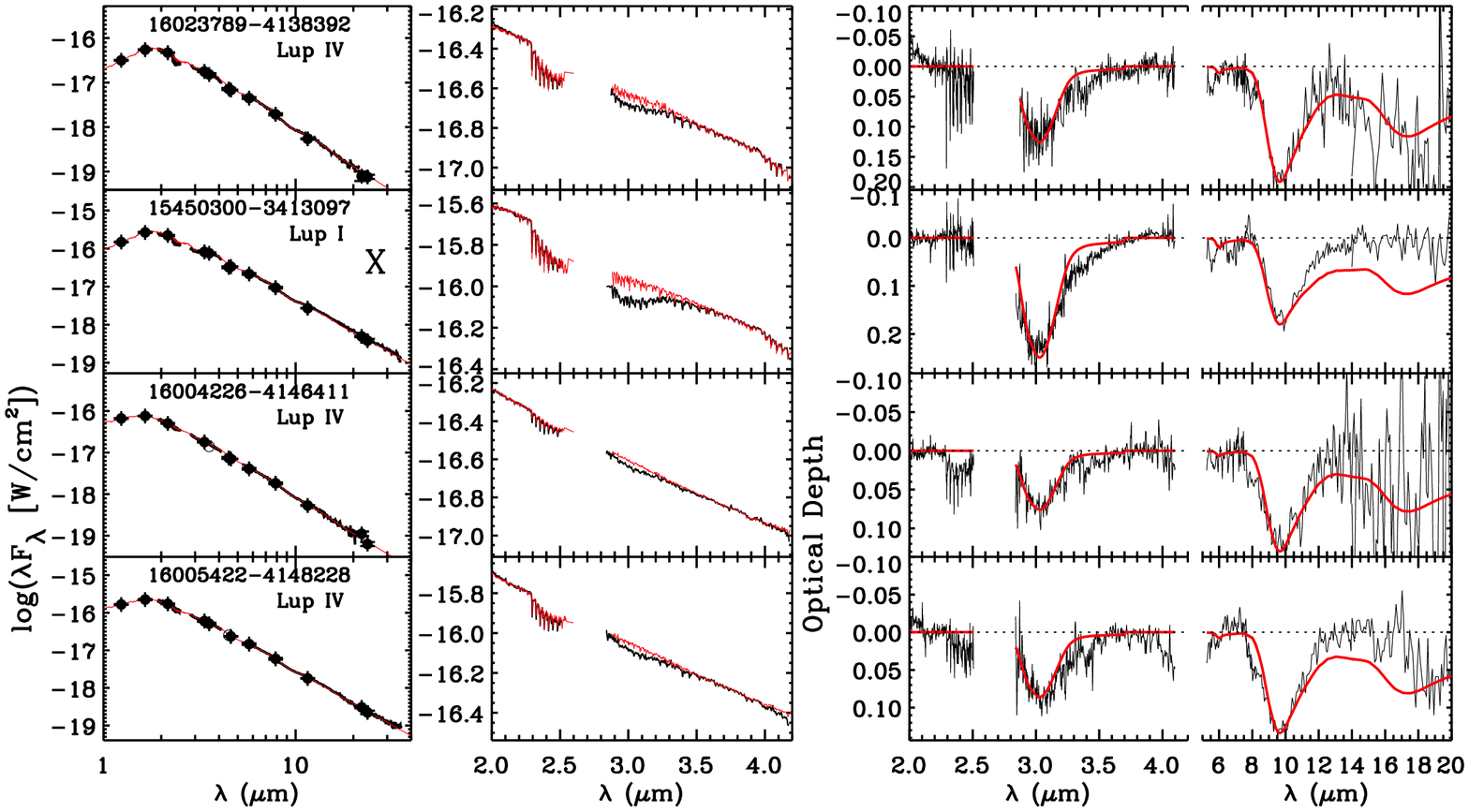}
\includegraphics[angle=90, scale=0.75]{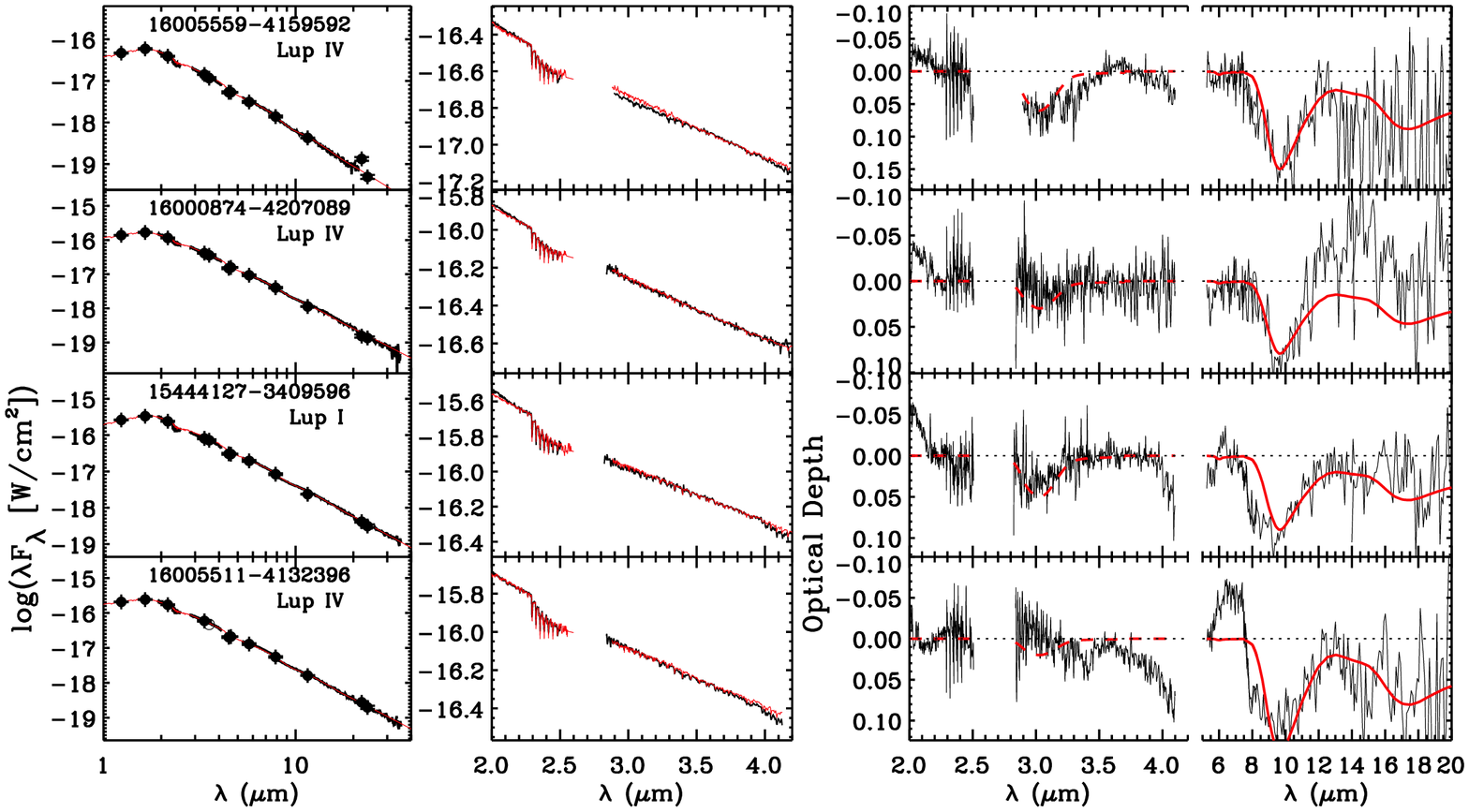}
\caption{(Continuation)}~\label{f:obs3}
\end{figure*}

\setcounter{figure}{1}

\begin{figure*}
\includegraphics[angle=90, scale=0.75]{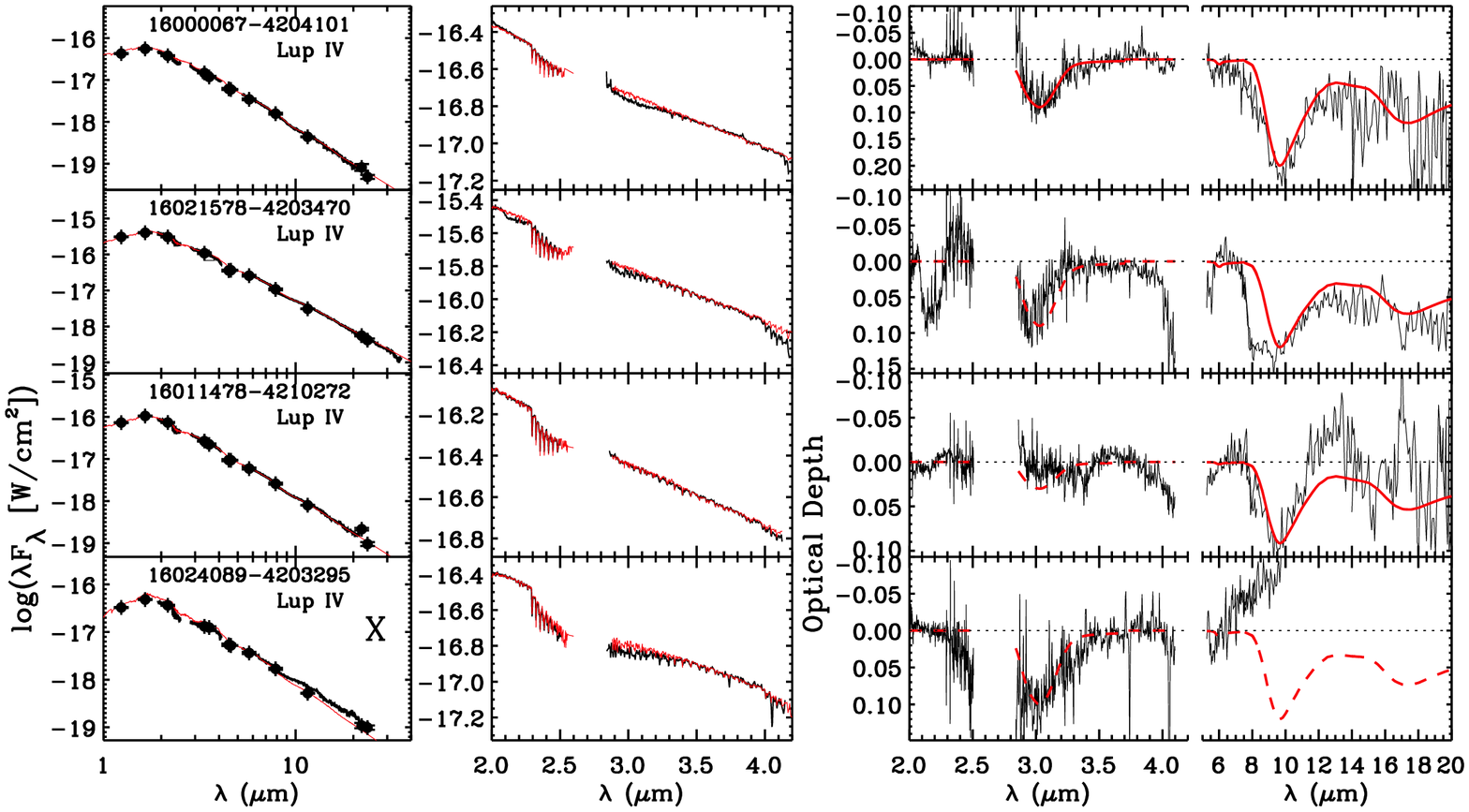}
\includegraphics[angle=90, scale=0.75]{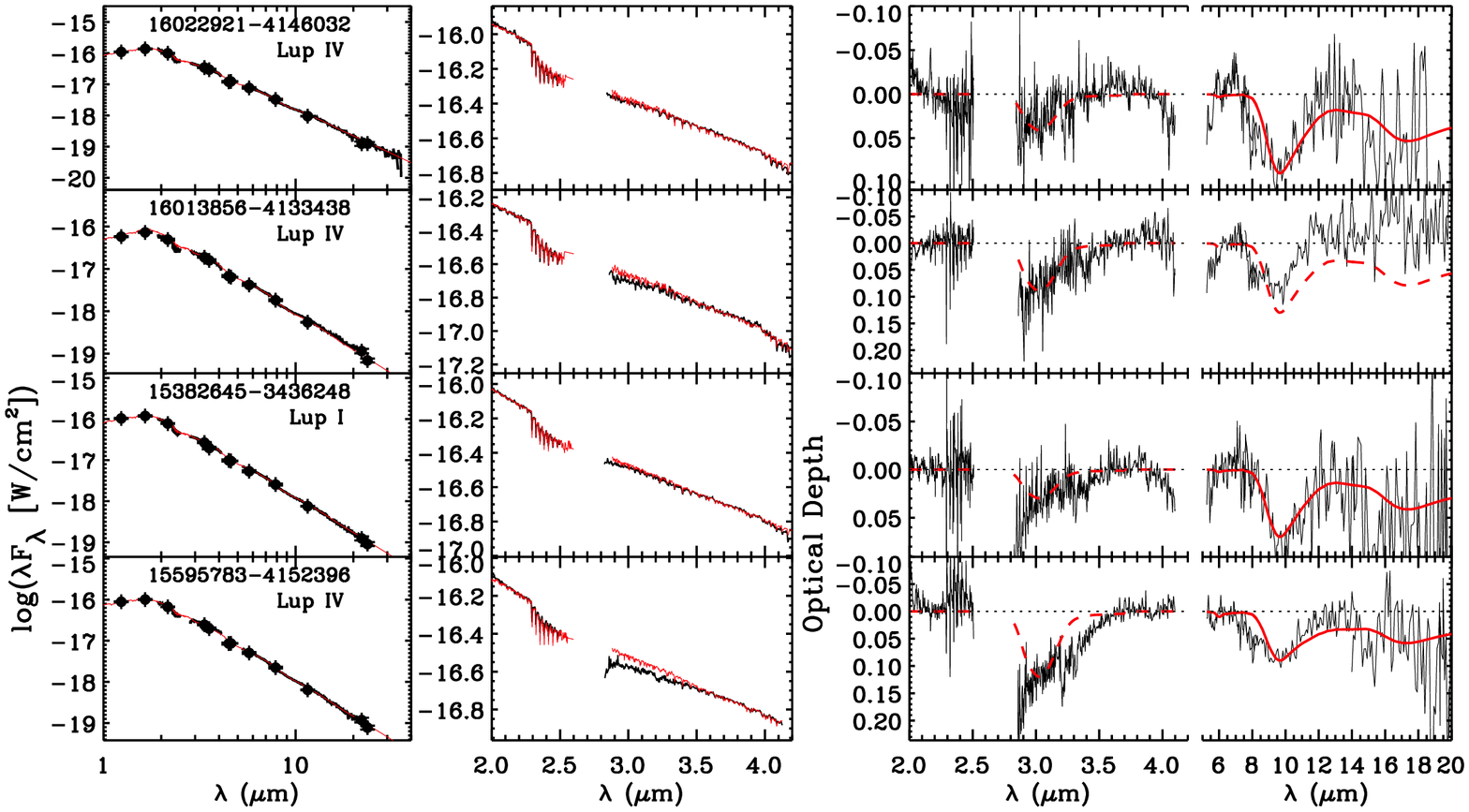}
\caption{(Continuation)}~\label{f:obs3}
\end{figure*}

\subsection{Continuum Determination}~\label{sec:cont}

\begin{figure}[h]
\includegraphics[angle=90, scale=0.55]{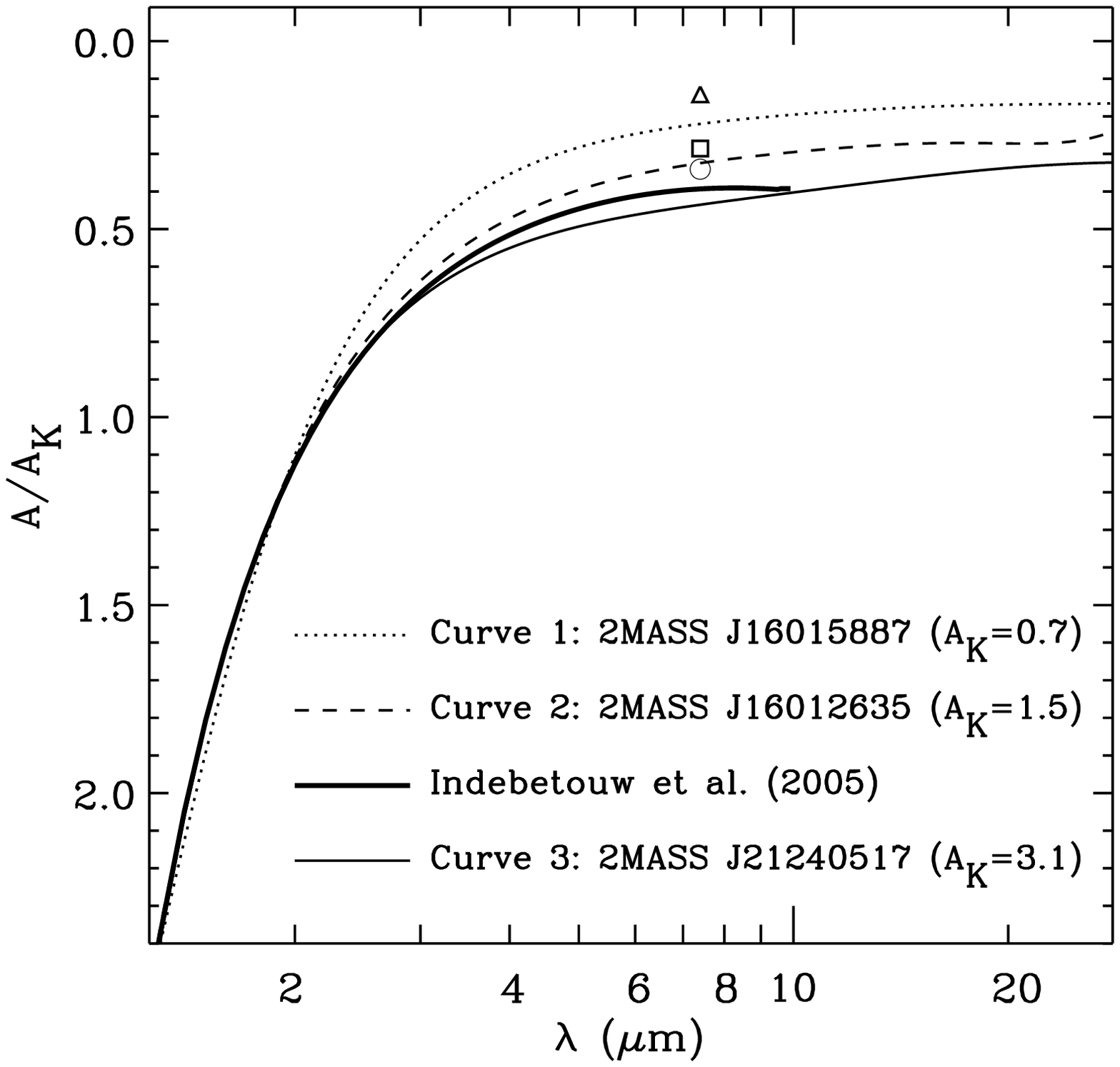}
\caption{Empirically derived, feature-free extinction curves used in
  the continuum fitting. The source of each curve is indicated in the
  plot. The curve for 2MASS J21240517 was derived in \citet{boo11}. The
  curve from \citet{ind05} is shown for comparison. It was derived
  from broad-band photometry and includes absorption by ice and
  dust features. The triangle, square, and circle represent the
  extinction at 7.4 \mum\ for $R_{\rm V}$=3.1, 4.0, and 5.5 models
  \citep{wei01}, respectively.}~\label{f:excur}
\end{figure}

\begin{figure}[h]
\includegraphics[angle=90, scale=0.50]{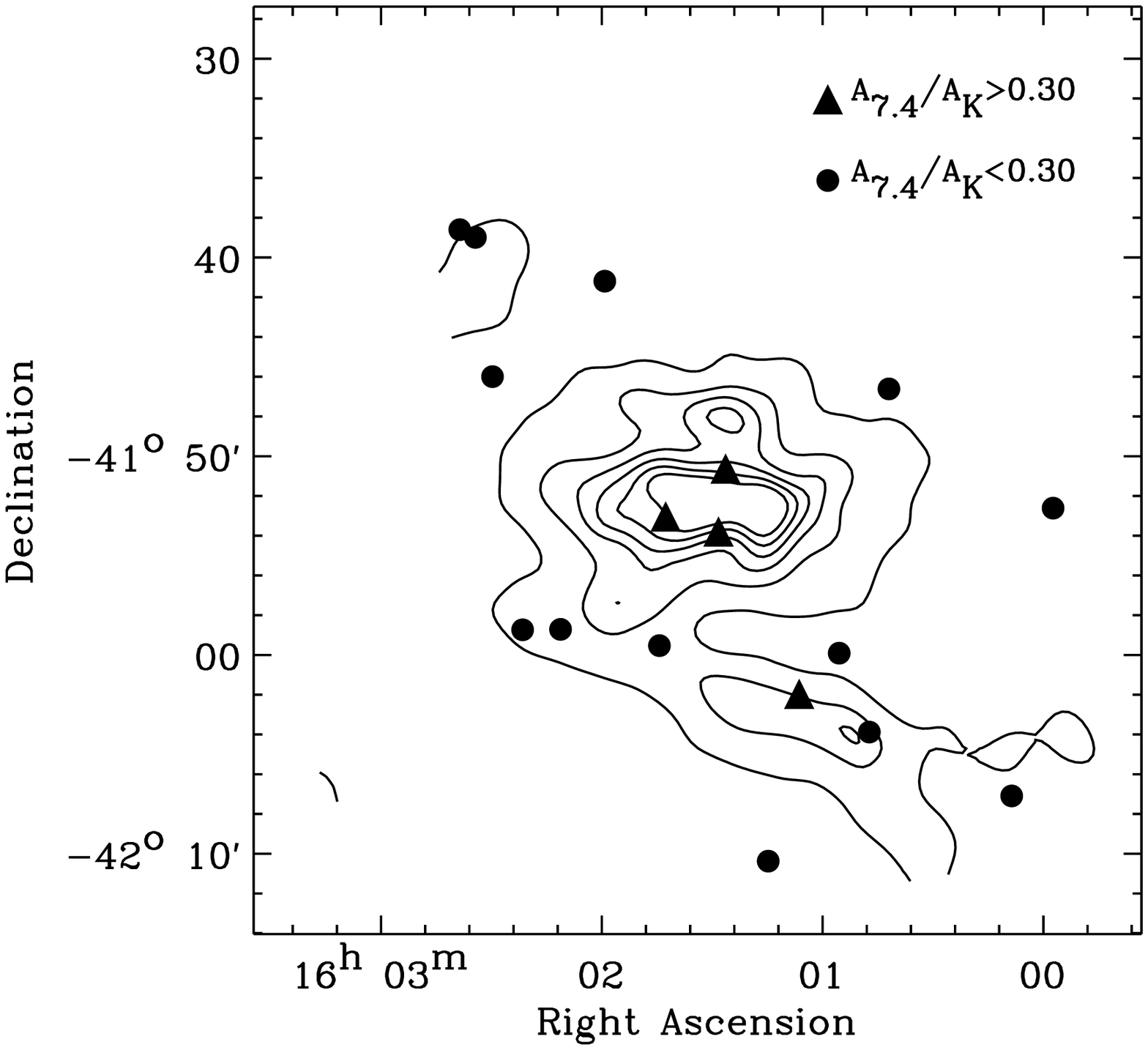}
\caption{Background stars with a continuum extinction ratio
  $A_{7.4}/A_{\rm K}>0.30$ (triangles) and $<0.30$ (bullets) plotted
  on top of the Lupus IV extinction map (same contours as in
  Fig.~\ref{f:maps}; \citealt{eva07}).}~\label{f:exlup4}
\end{figure}

The continua for the interstellar ice and dust absorption features
were determined in two steps. First, all available photometry and
spectra in the 1-4.2 \mum\ wavelength range were fitted with the full
IRTF database of observed stellar spectra \citep{ray09}, reddened
using the continuum extinction curves and H$_2$O ice model further
described below. These fits yield accurate values for the peak optical
depth of the 3.0 \mum\ band of H$_2$O ice ($\tau _{3.0}$), as the
continuum shape and photosperic absorption are corrected for
simultaneously.  Subsequently, the ground-based, WISE, and Spitzer
spectral and broad-band photometry over the full 1-30 \mum\ wavelength
range were fitted with thirteen model spectra of giants with spectral
types in the range G8 to M9 \citep{dec04, boo11}. These fits yield the
peak optical depth of the 9.7 \mum\ band of silicates, while $\tau
_{3.0}$ is fixed to the value found in the IRTF fits. Both fits yield
values for the extinction in the $K-$band ($A_{\rm K}$). Both the IRTF
and synthetic model fits use the same $\chi^2$ minimization routine
described in detail in \citet{boo11}, and have the same ingredients:

\begin{itemize}

\item {\it Feature-free, high resolution extinction curves.} Since it
  is the goal of this work to analyze the ice and dust absorption
  features, the IRTF database and synthetic stellar spectra must be
  reddened with a {\it feature-free} extinction curve. Such curve can
  be derived empirically, from the observed spectra themselves. The
  curve used in \citet{boo11} is derived for a high extinction line of
  sight ($A_{\rm K}=$3.10 mag) through the isolated core L1014. This
  curve does not always fit the lower extinction lines of sight
  through the Lupus clouds. Therefore, empirical, feature-free
  extinction curves are also derived from two Lupus IV sight-lines:
  2MASS J16012635-4150422 ($A_{\rm K}=$1.47) and 2MASS J16015887-4141159
  ($A_{\rm K}=$0.71). Throughout this paper, these will be referred to
  as extinction curves 1 ($A_{\rm K}=$0.71), 2 ($A_{\rm K}=$1.47), and
  3 ($A_{\rm K}=$3.10).  The three curves are compared in
  Fig.~\ref{f:excur}.  Clearly, lines of sight with lower $A_{\rm K}$
  values have lower mid-infrared continuum extinction. To compare the
  empirical curves with the models of \citet{wei01}, the median
  extinction in the 7.2-7.6 \mum\ range, relatively free of ice and
  dust absorption features, is calculated: $A_{7.4}/A_{\rm K}$=0.22,
  0.32, and 0.44, for curves 1-3 respectively. Curve 1 falls between
  the $R_{\rm V}=3.1$ ($A_{7.4}/A_{\rm K}$=0.14) and 4.0
  ($A_{7.4}/A_{\rm K}$=0.29) models, and thus corresponds to $R_{\rm
    V}\sim 3.5$. Curve 2 corresponds to $R_{\rm V}\sim 5.0$, and curve
  3 must have $R_{\rm V}$ well above 5.5 ($A_{7.4}/A_{\rm
    K}$=0.34). To further illustrate this point, $A_{7.4}/A_{\rm K}$
  is derived for all lines of sight and over-plotted on the extinction
  map of Lupus IV in Fig.~\ref{f:exlup4}. All lines of sight with
  $A_{\rm 7.4}$/$A_{\rm K}/>0.30$ lie near the high extinction peaks,
  while others lie in the low extinction outer regions.

\item {\it Laboratory H$_2$O ice spectra.} The optical constants of
  amorphous solid H$_2$O at $T=10$~K \citep{hud93} were used to
  calculate the absorption spectrum of ice spheres
  \citep{boh83}. Spheres with radii of 0.4 \mum\ fit the typical short
  wavelength profile and peak position of the observed 3 \mum\ bands
  best. While this may not be representative for actual dense cloud
  grain sizes and shapes, it suffices for fitting the H$_2$O band
  profiles and depths.

\item {\it Synthetic silicate spectra.} As for other dense cloud
  sight-lines and YSOs, the 9.7~\mum\ silicate spectra in the Lupus
  clouds are wider than those in the diffuse ISM \citep{bre11, boo11}.
  No evidence is found for narrower, diffuse medium type silicate
  bands.  Thus, the same synthetic silicate spectrum is used as in
  \citet{boo11}, i.e., for grains small compared to the wavelength,
  having a pyroxene to olivine optical depth ratio of 0.62 at the 9.7
  \mum\ peak.

\end{itemize}

The results of the continuum fitting are listed in Table~\ref{t:fits},
and the fits are plotted in Fig.~\ref{f:obs1} (red lines).  Two
reduced $\chi ^2$ values are given: one tracing the fit quality in the
1-4.2 \mum\ region using the IRTF database, and one tracing the longer
wavelengths using the synthetic stellar spectra. The IRTF fits were
done at a resolving power of $R=500$ and the reduced $\chi ^2$ values
are very sensitive to the fit quality of the photospheric CO overtone
lines at 2.25-2.60~\mum, as well as other photospheric lines, including
the onset of the SiO $\Delta v=2$ overtone band at 4.0 \mum.  The
wavelength region of 3.09-3.7 \mum\ is excluded in the $\chi ^2$
determination because the long wavelength wing of the H$_2$O ice band
is not part of the model. In some cases, the flux scale of the
$L-$band spectrum relative to the $K-$band had to be multiplied with a
scaling factor to obtain the most optimal fit. These adjustments are
attributed to the statistical uncertainties in the broad-band
photometry used to scale the observed spectra, i.e., they are
generally within 1$\sigma$ of the photometric error bars and at most
2.1$\sigma$ in four cases. Finally, the fits were inspected and $\tau
_{3.0}$ values were converted to 3$\sigma$ upper limits in case no
distinct 3.0 \mum\ ice band was present, but rather a shallow, broader
residual (dashed lines in Fig.~\ref{f:obs1}).

\begin{deluxetable*}{lcccccccc}
\tabletypesize{\scriptsize}
\tablecolumns{9}
\tablewidth{0pc}
\tablecaption{Continuum Fit Parameters~\label{t:fits}}
\tablehead{
\colhead{Source}& \multicolumn{2}{c}{Spectral Type}      &\colhead{$A_{\rm K   }  $\tablenotemark{c}}&
                  \colhead{$\tau_{3.0}$\tablenotemark{d}} &\colhead{$\tau_{\rm 9.7}$\tablenotemark{e}}&
                  \colhead{Ext. Curve\tablenotemark{f}}  &
                  \multicolumn{2}{c}{$\chi_{\nu}^2$}\\
\colhead{2MASS~J}&\colhead{Model\tablenotemark{a}} &\colhead{IRTF\tablenotemark{b}} & 
                  \colhead{mag            } &\colhead{              } & 
                  \colhead{               } &\colhead{              } &
                  \colhead{Model\tablenotemark{g}          } & \colhead{IRTF\tablenotemark{h}         }\\}
\startdata
$16014254-4153064$ & M1 (M0-M3) & HD120052/M2 & 2.46       (0.10)  & 1.03 (0.04)  & 0.55 (0.03)  & 2 & 0.61 & 0.78 \\
$16004739-4203573$ & K5 (K2-K7) & HD132935/K2 & 2.03       (0.08)  & 0.63 (0.06)  & 0.48 (0.03)  & 2 & 1.56 & 0.76 \\
$16010642-4202023$ & K0 (G8-K0) & HD135722/G8 & 1.91       (0.07)  & 0.74 (0.06)  & 0.48 (0.04)  & 2 & 0.35 & 0.78 \\
$15423699-3407362$ & G8 (G8-K0) & HD222093/G9 & 1.79       (0.09)  & 0.83 (0.04)  & 0.42 (0.02)  & 3 & 0.42 & 0.35 \\
$16012635-4150422$ & M0 (M0-M1) & HD204724/M1 & 1.65       (0.08)  & 0.71 (0.04)  & 0.35 (0.04)  & 2 & 1.22 & 0.60 \\
$15452747-3425184$ & M1 (M0-M1) & HD204724/M1 & 1.58       (0.10)  & 0.63 (0.02)  & 0.39 (0.02)  & 2 & 29.98 & 0.37 \\
$16012825-4153521$ & K7 (K3-K7) & HD35620/K3.5 & 1.57       (0.12)  & 0.53 (0.05)  & 0.34 (0.03)  & 3 & 0.52 & 0.44 \\
$16004925-4150320$ & M1 (M0-M1) & HD204724/M1 & 0.97       (0.05)  & 0.26 (0.06)  & 0.26 (0.03)  & 2 & 2.05$^{\rm i}$ & 0.38 \\
$16021102-4158468$ & K5 (K2-K7) & HD132935/K2 & 0.72       (0.05)  & 0.11 (0.02)  & 0.25 (0.03)  & 1 & 1.20 & 0.41 \\
$16014426-4159364$ & K5 (K2-K7) & HD132935/K2 & 0.67       (0.03)  & 0.16 (0.03)  & 0.19 (0.03)  & 1 & 0.66 & 0.45 \\
$16022128-4158478$ & M1 (M0-M2) & HD120052/M2 & 0.66       (0.05)  & 0.27 (0.02)  & 0.21 (0.02)  & 1 & 0.61 & 0.23 \\
$16015887-4141159$ & M1 (M0-M1) & HD204724/M1 & 0.59       (0.05)  & 0.16 (0.05)  & 0.20 (0.02)  & 1 & 14.34 & 0.92 \\
$16023370-4139027$ & M0 (M0-M2) & HD120052/M2 & 0.57       (0.06)  & 0.17 (0.04)  & 0.28 (0.04)  & 1 & 1.06 & 0.35 \\
$16023789-4138392$ & M1 (M0-M3) & HD219734/M2.5 & 0.53       (0.05)  & 0.13 (0.03)  & 0.19 (0.03)  & 1 & 0.67 & 0.41 \\
$16004226-4146411$ & G8 (G8-K1) & HD25975/K1 & 0.46       (0.02)  & 0.08 (0.02)  & 0.13 (0.01)  & 2 & 0.48 & 0.84 \\
$16005422-4148228$ & K5 (K2-K7) & HD132935/K2 & 0.45       (0.04)  & 0.09 (0.02)  & 0.13 (0.02)  & 1 & 1.51$^{\rm i}$ & 0.32 \\
$16000067-4204101$ & K3 (K0-K4) & HD2901/K2 & 0.41       (0.03)  & 0.11 (0.03)  & 0.20 (0.02)  & 1 & 1.50 & 0.66 \\
$16021578-4203470$ & M1 (K7-M1) & HD204724/M1 & 0.36       (0.06)  & $<$0.09 & 0.12 (0.03)  & 1 & 11.28$^{\rm i}$ & 1.88 \\
$16011478-4210272$ & M0 (K5-M0) & HD120477/K5.5 & 0.34       (0.04)  & $<$0.03 & 0.09 (0.05)  & 1 & 1.53 & 0.26 \\
$16005559-4159592$ & K7 (K3-K7) & HD35620/K3.5 & 0.31       (0.05)  & $<$0.06 & 0.15 (0.03)  & 1 & 0.78 & 0.74 \\
$16000874-4207089$ & K4 (K0-K4) & HD132935/K2 & 0.29       (0.05)  & $<$0.03 & 0.08 (0.02)  & 1 & 0.77 & 0.43 \\
$15444127-3409596$ & M0 (M0-M1) & HD213893/M0 & 0.26       (0.05)  & $<$0.05 & 0.09 (0.02)  & 1 & 2.52$^{\rm i}$ & 0.31 \\
$16005511-4132396$ & K5 (K5-M1) & HD120477/K5.5 & 0.23       (0.07)  & $<$0.02 & 0.14 (0.04)  & 1 & 1.17$^{\rm i}$ & 1.09 \\
$16022921-4146032$ & M1 (M0-M2) & HD120052/M2 & 0.17       (0.04)  & $<$0.04 & 0.09 (0.03)  & 1 & 0.63 & 0.40 \\
$16013856-4133438$ & M1 (M1-M5) & HD219734/M2.5 & 0.16       (0.06)  & $<$0.09 & $<$0.13 & 1 & 2.81$^{\rm i}$ & 0.34 \\
$15595783-4152396$ & K7 (K5-K7) & HD120477/K5.5 & 0.14       (0.03)  & $<$0.12 & 0.09 (0.02)  & 1 & 1.41 & 0.93 \\
$15382645-3436248$ & M1 (K7-M1) & HD213893/M0 & 0.14       (0.05)  & $<$0.03 & 0.07 (0.03)  & 1 & 0.38 & 0.36 \\
$15424030-3413428$ & M3 (M2-M6) & HD27598/M4 & 0.77       (0.04)  & 0.36 (0.05)  & $<$0.30 & 1 & 21.01$^{\rm j}$ & 0.21 \\
$15425292-3413521$ & M6 (M1-M6) & HD27598/M4 & 0.70       (0.06)  & 0.36 (0.03)  & $<$0.25 & 1 & 44.36$^{\rm j}$ & 1.42 \\
$16003535-4209337$ & M1 (M0-M4) & HD28487/M3.5 & 0.60       (0.09)  & 0.18 (0.04)  & 0.25 (0.09)  & 1 & 3.79$^{\rm j}$ & 0.99 \\
$15450300-3413097$ & M1 (M0-M4) & HD28487/M3.5 & 0.47       (0.11)  & 0.25 (0.04)  & 0.18 (0.04)  & 1 & 2.07$^{\rm j}$ & 0.22 \\
$16024089-4203295$ & M6 (M2-M6) & HD27598/M4 & 0.31       (0.03)  & $<$0.10 & $<$0.12 & 1 & 12.03$^{\rm j}$ & 0.65 \\
\enddata
\tablenotetext{a}{ Best fitting spectral type using the synthetic
  models listed in Table~2 of \citet{boo11} over the full observed
  wavelength range. For all spectral types the luminosity class is
  III. The uncertainty range is given in parentheses.}
\tablenotetext{b}{ Best fitting 1-4 \mum\ spectrum from the IRTF
  database of \citet{ray09}. All listed spectral types have luminosity
  class III.}
\tablenotetext{c}{ Extinction in the 2MASS $K$-band.}
\tablenotetext{d}{ Peak absorption optical depth of the 3.0~\mum\ H$_2$O ice band.}
\tablenotetext{e}{ Peak absorption optical depth of the 9.7~\mum\ band of silicates.}
\tablenotetext{f}{Extinction curve used: 
  1-derived from 2MASS J16015887-4141159 (Lupus IV, $A_{\rm K}$=0.71),
  2-derived from 2MASS J16012635-4150422 (Lupus IV, $A_{\rm K}$=1.47),
  3-derived from 2MASS J21240517+4959100 (L1014, $A_{\rm K}$=3.10; \citealt{boo11})}
\tablenotetext{g}{ Reduced $\chi^2$ values of the model spectrum with
  respect to the observed spectral data points in the 5.2-5.67 and
  7.2-14 \mum\ wavelength ranges. Values higher than 1.0 generally
  indicate that the model underestimates the bands of photospheric CO
  at 5.3 \mum\ and SiO at 8.0 \mum. In the following cases,
  $\chi_{\nu}^2$ values are high for different reasons.  2MASS
  J15452747-3425184: very small error bars, fit is excellent for
  purpose of this work.  2MASS J15424030-341342: shows PAH emission
  bands and has shallower slope than model.  2MASS J15425292-3413521:
  offset and shallower slope than model.  2MASS J16024089-4203295,
  2MASS J16013856-4133438, 2MASS J16003535-4209337, and 2MASS
  J15450300-3413097: shallower slope than model.  2MASS
  J16023370-4139027: steeper slope than model.}
\tablenotetext{h}{ Reduced $\chi^2$ values of the IRTF spectra to all
  observed near-infrared photometry and spectra ($J$, $H$, $K$, and
  $L$-band), excluding the long-wavelength wing of the 3.0~\mum\ ice
  band.}
\tablenotetext{i}{ A poor fit to the photospheric CO and SiO regions
  near 5.3 and 8.0 \mum\ prohibits the analysis of the interstellar
  5-8 \mum\ ice and 9.7 \mum\ silicate features for this source.}
\tablenotetext{j}{ The model systematically underestimates the
  emission at longer wavelengths, and this source is not considered a
  bona fide background star in the analysis.}
\end{deluxetable*}

While generally excellent fits are obtained with the IRTF database,
this is not always the case at longer wavelengths with the synthetic
spectra. The reduced $\chi ^2$ values (Table~\ref{t:fits}) were
determined in the 5.3-5.67 and 7.2-14 \mum\ wavelength regions, which
do not only cover the interstellar 9.7 \mum\ silicate and the 13
\mum\ H$_2$O libration ice band, but also the broad photospheric CO
($\sim$5.3 \mum) and SiO (8.0 \mum) bands.  Inspection of the best
fits shows that reduced $\chi ^2$ values larger than 1.0 generally
indicate deviations in the regions of the photospheric bands, even if
the near-infrared CO overtone lines are well matched. For this reason,
six sources (labeled in Table~\ref{t:fits}) were excluded from a
quantitative analysis of the 5-7 and 9.7 \mum\ interstellar absorption
bands. Other causes for high reduced $\chi ^2$ values for some
sight-lines are further explained in the footnotes of
Table~\ref{t:fits}. Notably, for five sight-lines, a systematic
continuum excess is observed. One of these is a Class III YSO
(\S\ref{sec:sou}).  These five sources will not be treated as
background stars further on.  In general, however, a good agreement
was found between the IRTF and synthetic spectra fits: all best-fit
IRTF models are of luminosity class III (justifying the use of the
synthetic spectra of giants), the spectral types agree to within 3
sub-types, and the $A_{\rm K}$ values agree within the uncertainties.


\subsection{Ice Absorption Band Strengths and Abundances}~\label{sec:col}

All detected absorption features are attributed to interstellar ices,
except the 9.7 \mum\ band of silicates. Their strengths are determined
here and converted to column densities and abundances
(Tables~\ref{t:colden} and \ref{t:tau}), using the intrinsic
integrated band strengths summarized in \citet{boo11}.  Uncertainties
are at the 1$\sigma$ level, and upper limits are of 3$\sigma$
significance.

\subsubsection{H$_2$O}~\label{sec:h2o}

\begin{turnpage}
\begin{deluxetable*}{llccrrrrrr}
\tabletypesize{\footnotesize}
\setlength{\tabcolsep}{0.06in} 
\tablecolumns{10}
\tablewidth{0pc}
\tablecaption{Ice Column Densities and Abundances~\label{t:colden}}
\tablehead{
\colhead{Source}& \colhead{$N$(H$_2$O)\tablenotemark{a}}     & 
                  \colhead{$N_{\rm H}$\tablenotemark{b}}   &
                  \colhead{$x$(H$_2$O)\tablenotemark{c}}   &
                  \multicolumn{2}{c}{$N$(NH$_4^+$)\tablenotemark{d, e, f}}   &
                  \multicolumn{2}{c}{$N$(CO$_2$)\tablenotemark{f}} &
                  \multicolumn{2}{c}{$N$(CO)\tablenotemark{f}} \\
\colhead{2MASS~J}& \colhead{$10^{18}$ } & 
                   \colhead{$10^{22}$ } & 
                   \colhead{$10^{-5}$ } & 
                   \colhead{$10^{17}$ } & 
                   \colhead{      } &
                   \colhead{$10^{17}$ } & 
                   \colhead{      } &
                   \colhead{$10^{17}$ } & 
                   \colhead{      } \\
\colhead{       }& \colhead{ \sqcm} & 
                   \colhead{ \sqcm} & 
                   \colhead{      } & 
                   \colhead{      } & 
                   \colhead{\%H$_2$O      } &
                   \colhead{ \sqcm} & 
                   \colhead{\%H$_2$O      } &
                   \colhead{ \sqcm} & 
                   \colhead{\%H$_2$O      } \\}
\startdata
\multicolumn{10}{c}{Background stars}\\
\hline
  $16014254-4153064        $ &       1.73                  (0.18) &  7.56 ( 0.30) &     2.29 (0.26) &     1.24 (0.43) &     7.14 (2.60) &     5.58 (2.41) &    32.14 (14.3) &     6.81 (0.30) &    43.10 (5.02) \\
  $16004739-4203573        $ &       1.06                  (0.14) &  6.25 ( 0.26) &     1.70 (0.23) &  $<$1.15        & $<$12.56        &     7.64 (3.31) &    71.88 (32.6) &  \nodata        &  \nodata        \\
  $16010642-4202023        $ &       1.24                  (0.15) &  5.89 ( 0.23) &     2.12 (0.28) &  $<$1.42        & $<$13.07        &    12.46 (4.27) &    99.80 (36.5) &  \nodata        &  \nodata        \\
  $15423699-3407362        $ &       1.40                  (0.15) &  5.52 ( 0.29) &     2.53 (0.31) &     0.84 (0.31) &     6.01 (2.37) &  $<$4.02        & $<$32.34        &  \nodata        &  \nodata        \\
  $16012635-4150422        $ &       1.19                  (0.13) &  5.07 ( 0.24) &     2.36 (0.29) &     1.26 (0.45) &    10.58 (3.96) &     5.76 (2.81) & $<$72.46        &  \nodata        &  \nodata        \\
  $15452747-3425184        $ &       1.06                  (0.11) &  4.86 ( 0.30) &     2.18 (0.26) &     0.26 (0.05) &     2.50 (0.57) &     4.54 (0.28) &    42.74 (5.21) &     2.45 (0.08) &    25.44 (2.78) \\
  $16012825-4153521        $ &       0.89                  (0.12) &  4.85 ( 0.38) &     1.84 (0.29) &  $<$0.96        & $<$12.52        &     7.18 (2.91) &    80.32 (34.5) &  \nodata        &  \nodata        \\
  $16004925-4150320        $ &       0.43                  (0.11) &  2.97 ( 0.16) &     1.47 (0.37) &  \nodata        &  \nodata        &     1.23 (0.52) &    28.08 (13.8) &  $<$1.05        & $<$35.36        \\
  $16021102-4158468        $ &       0.18                  (0.04) &  2.21 ( 0.16) &     0.84 (0.19) &  $<$0.95        & $<$65.92        &     3.48 (1.06) &    187.8 (71.2) &  $<$0.62        & $<$47.73        \\
  $16014426-4159364        $ &       0.27                  (0.06) &  2.05 ( 0.10) &     1.31 (0.30) &  $<$1.01        & $<$48.32        &  $<$2.82        & $<$134.9        &  \nodata        &  \nodata        \\
  $16022128-4158478        $ &       0.45                  (0.05) &  2.05 ( 0.14) &     2.22 (0.32) &  $<$0.88        & $<$22.24        &     2.68 (1.15) &    58.87 (26.4) &  \nodata        &  \nodata        \\
  $16015887-4141159        $ &       0.27                  (0.08) &  1.83 ( 0.14) &     1.47 (0.47) &  $<$0.51        & $<$27.63        &  $<$0.60        & $<$32.71        &  \nodata        &  \nodata        \\
  $16023370-4139027        $ &       0.28                  (0.07) &  1.76 ( 0.19) &     1.62 (0.46) &  $<$0.97        & $<$46.66        &     1.33 (0.57) & $<$70.63        &  \nodata        &  \nodata        \\
  $16023789-4138392        $ &       0.21                  (0.05) &  1.62 ( 0.16) &     1.35 (0.38) &  $<$1.18        & $<$74.12        &  $<$2.42        & $<$151.3        &  \nodata        &  \nodata        \\
  $16004226-4146411        $ &       0.13                  (0.04) &  1.43 ( 0.06) &     0.94 (0.30) &  $<$0.89        & $<$96.51        &  $<$3.90        & $<$422.8        &  \nodata        &  \nodata        \\
  $16005422-4148228        $ &       0.15                  (0.03) &  1.39 ( 0.12) &     1.09 (0.26) &  \nodata        &  \nodata        &     0.86 (0.39) & $<$87.75        &  \nodata        &  \nodata        \\
  $16000067-4204101        $ &       0.18                  (0.04) &  1.28 ( 0.10) &     1.45 (0.40) &  $<$1.02        & $<$75.31        &  $<$3.34        & $<$246.4        &  \nodata        &  \nodata        \\
  $16021578-4203470        $ &    $<$0.15                         &  1.12 ( 0.18) &  $<$1.61        &  \nodata        &  \nodata        &  $<$0.83        &  \nodata        &  \nodata        &  \nodata        \\
  $16011478-4210272        $ &    $<$0.05                         &  1.03 ( 0.14) &  $<$0.56        &  $<$0.63        &  \nodata        &     1.33 (0.61) &  \nodata        &  \nodata        &  \nodata        \\
  $16005559-4159592        $ &    $<$0.10                         &  0.96 ( 0.14) &  $<$1.22        &  $<$1.10        &  \nodata        &  $<$3.56        &  \nodata        &  \nodata        &  \nodata        \\
  $16000874-4207089        $ &    $<$0.05                         &  0.89 ( 0.15) &  $<$0.68        &  $<$0.78        &  \nodata        &  $<$1.72        &  \nodata        &  \nodata        &  \nodata        \\
  $15444127-3409596        $ &    $<$0.08                         &  0.82 ( 0.15) &  $<$1.27        &  \nodata        &  \nodata        &  $<$1.12        &  \nodata        &  \nodata        &  \nodata        \\
  $16005511-4132396        $ &    $<$0.03                         &  0.71 ( 0.21) &  $<$0.67        &  \nodata        &  \nodata        &  $<$1.42        &  \nodata        &  \nodata        &  \nodata        \\
  $16022921-4146032        $ &    $<$0.06                         &  0.51 ( 0.11) &  $<$1.67        &  $<$0.83        &  \nodata        &     1.72 (0.68) &  \nodata        &  \nodata        &  \nodata        \\
  $16013856-4133438        $ &    $<$0.15                         &  0.50 ( 0.18) &  $<$4.73        &  \nodata        &  \nodata        &  $<$2.27        &  \nodata        &  \nodata        &  \nodata        \\
  $15382645-3436248        $ &    $<$0.05                         &  0.44 ( 0.15) &  $<$1.71        &  $<$0.85        &  \nodata        &  $<$3.30        &  \nodata        &  \nodata        &  \nodata        \\
  $15595783-4152396        $ &    $<$0.20                         &  0.44 ( 0.09) &  $<$5.90        &  $<$0.73        &  \nodata        &  $<$3.06        &  \nodata        &  \nodata        &  \nodata        \\
\hline
\multicolumn{10}{c}{Sources with long-wavelength excess}\\
\hline
  $15424030-3413428        $ &       0.60                  (0.10) &  2.38 ( 0.12) &     2.55 (0.47) &  \nodata        &  \nodata        &     3.23 (0.50) &    53.19 (12.6) &  \nodata        &  \nodata        \\
  $15425292-3413521        $ &       0.60                  (0.08) &  2.16 ( 0.18) &     2.81 (0.44) &  \nodata        &  \nodata        &     3.85 (0.69) &    63.38 (14.2) &  $<$0.63        & $<$13.24        \\
  $16003535-4209337        $ &       0.30                  (0.08) &  1.86 ( 0.27) &     1.63 (0.50) &  \nodata        &  \nodata        &  $<$4.26        & $<$192.3        &  \nodata        &  \nodata        \\
  $15450300-3413097        $ &       0.42                  (0.07) &  1.44 ( 0.33) &     2.93 (0.84) &  \nodata        &  \nodata        &     0.89 (0.36) &    21.27 (9.44) &  $<$0.79        & $<$25.04        \\
  $16024089-4203295        $ &    $<$0.16                         &  0.97 ( 0.09) &  $<$1.91        &  \nodata        &  \nodata        &  $<$2.64        &  \nodata        &  \nodata        &  \nodata        \\
\hline
\multicolumn{10}{c}{Embedded YSO}\\
\hline
  $15430131-3409153^{\rm g}$ &       14.8                  (4.0)  & 39.33 ( 3.91) &     3.76 (1.08) &     5.77 (0.44) &      3.9  (0.3) &     51.8 (5.92) &       35    (4) &  \nodata\nodata &  \nodata\nodata \\
\enddata
\tablecomments{The sources are sorted in order of decreasing $A_{\rm
    K}$ values (Table~\ref{t:fits}). Column densities were determined
  using the intrinsic integrated band strengths summarized in
  \citet{boo08}. Uncertainties (1$\sigma$) are indicated in brackets
  and upper limits are of 3$\sigma$ significance. The species
  CH$_3$OH, H$_2$CO, HCOOH, CH$_4$, and NH$_3$ are not listed in this
  table, but their upper limits are discussed in \S\ref{sec:ch3oh} and
  \S\ref{sec:hcooh}. }
\tablenotetext{a}{An uncertainty of 10\% in the intrinsic integrated
  band strength is taken into account in the listed column density
  uncertainties.}
\tablenotetext{b}{Column density of HI and H$_2$, calculated from
  $A_{\rm K}$ (see \S\ref{sec:complex} for details).}
\tablenotetext{c}{Solid H$_2$O abundance with respect to $N_{\rm H}$.}
\tablenotetext{d}{Assuming that the entire 6.4-7.2 \mum\ region, after
  H$_2$O subtraction, is due to NH$_4^+$. }
\tablenotetext{e}{No values are given for sources with large
  photospheric residuals.}
\tablenotetext{f}{Column densities with significance $<2\sigma$ were
  converted to $3\sigma$ upper limits.}
\tablenotetext{g}{The YSO IRAS 15398-3359. Ice column densities were
  taken from \citet{boo08} and \citet{pon08}. Not listed are
  detections of CH$_4$ ($6\pm 2$\%; \citealt{obe08}), NH$_3$ ($7.6\pm
  1.7$\%; \citealt{bot10}), CH$_3$OH ($10.3\pm 0.8$\%;
  \citealt{boo08,bot10}), and HCOOH ($1.9\pm 0.2$\%;
  \citealt{boo08}). $N_{\rm H}$ was calculated from $\tau_{9.7}=3.32\pm 0.33$
  \citep{boo08} and Eqs.~\ref{eq:tau97ak} and~\ref{eq:nh2}.}
\end{deluxetable*}
\end{turnpage}

The peak optical depths of the 3.0 \mum\ H$_2$O stretching mode listed
in Table~\ref{t:fits} were converted to H$_2$O column densities
(Table~\ref{t:colden}), by integrating the H$_2$O model spectra
(\S\ref{sec:cont}) over the 2.7-3.4 \mum\ range.  An uncertainty of
10\% in the intrinsic integrated band strength is taken into account
in the listed column density uncertainties. Subsequently, H$_2$O
abundances relative to $N_{\rm H}$, the total hydrogen (HI and H$_2$)
column density along the line of sight, were derived. $N_{\rm H}$ was
calculated from the $A_{\rm K}$ values of Table~\ref{t:fits},
following the Oph cloud relation of \citet{boh78}:

\begin{equation}
N_{\rm H}=15.4 \times 10^{21} \times (A_{\rm V}/A_{\rm K})/R_{\rm V} \times
A_{\rm K}~[{\rm cm}^{-2}] ~\label{eq:nh1}
\end{equation}

\noindent Here, $R_{\rm V}$=4.0 and $A_{\rm V}/A_{\rm K}$=8.0
\citep{car89} are taken for the Lupus clouds, which gives

\begin{equation}
N_{\rm H}=3.08 \times 10^{22} \times A_{\rm K}~[{\rm cm}^{-2}]
~\label{eq:nh2}
\end{equation}

\noindent The resulting H$_2$O abundances are typically few$\times
10^{-5}$ (Table~\ref{t:colden}). The uncertainty in Eq.~\ref{eq:nh2}
is not included in Table~\ref{t:colden}. This ``absolute'' uncertainty
is estimated to be on the order of 30\%, based on conversion factors
for $R_{\rm V}$ in the range of 3.5-5.5.

\subsubsection{5-7 \mum\ Bands}

\begin{figure}[b]
\includegraphics[angle=90, scale=0.65]{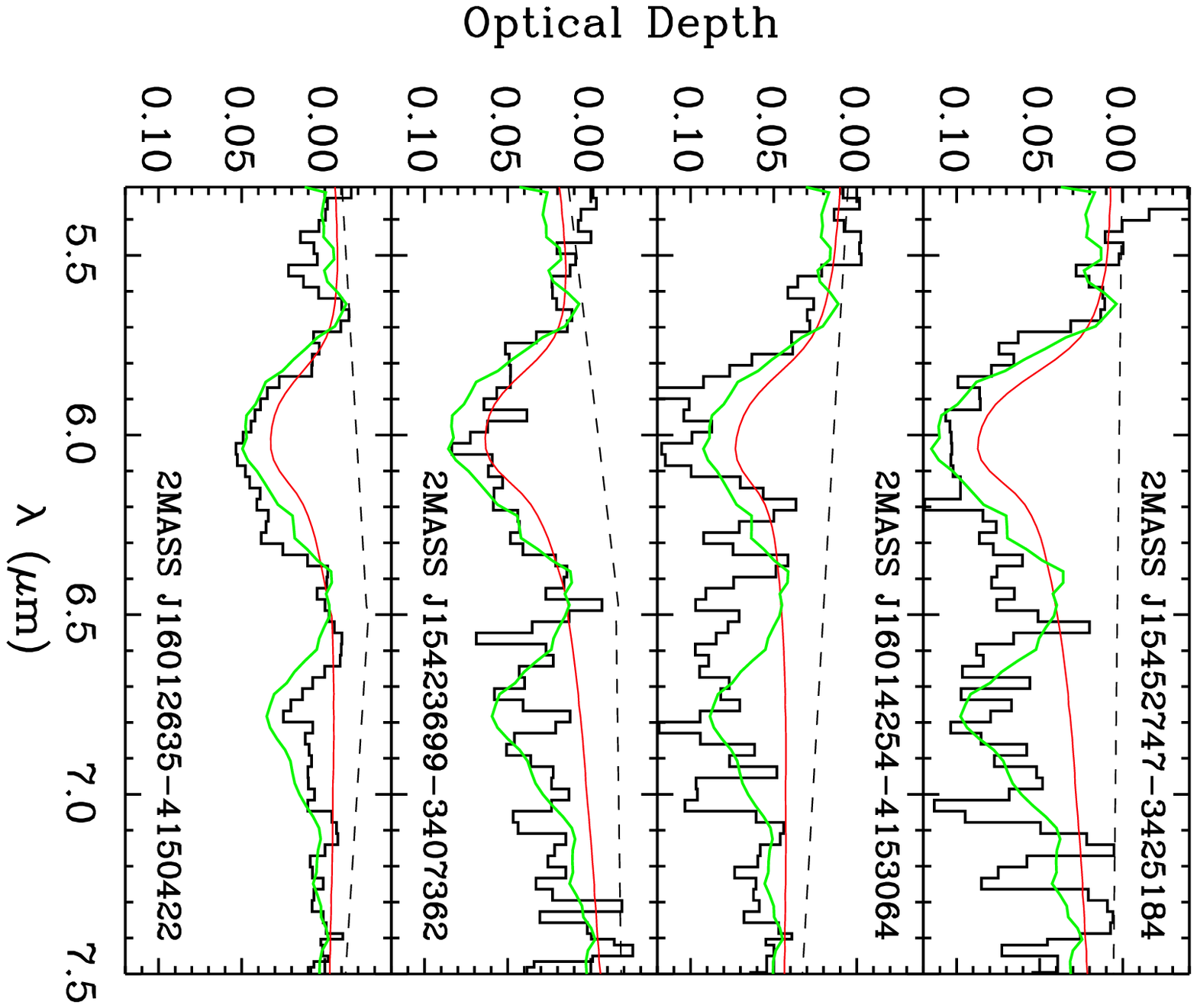}
\caption{Lines of sight with the most securely detected 6.0
  \mum\ bands, and also the only lines of sight in which the 6.85
  \mum\ band is detected at the $>3\sigma$ level. The red line
  represents the spectrum of solid H$_2$O at $T=10$~K at the column
  density derived from the 3.0 \mum\ O$-$H stretching mode. For
  comparison, the spectrum of the TMC background star Elias 16
  (2MASS J04393886+2611266; \citealt{kne05}) is overplotted
  (green). The dashed line is the local baseline adopted, in addition
  to the global continuum discussed in
  \S\ref{sec:cont}.}~\label{f:6_68um}
\end{figure}

The well known 5-7 \mum\ absorption bands have for the first time been
detected toward Lup I and IV background stars
(Fig.~\ref{f:6_68um}). Eight lines of sight show the 6.0 \mum\ band
and four the 6.85 \mum\ band. In particular for the latter, the
spectra are noisy and the integrated intensity is just at the
3$\sigma$ level in three sources. The line depths are in agreement
with other clouds, however, as can be seen by the green line in
Fig.~\ref{f:6_68um}, representing Elias 16 in the TMC.  The integrated
intensities and upper limits are listed in Table~\ref{t:tau}.  They
were derived after subtracting a local, linear baseline, needed
because the accuracy of the global baseline is limited to
$\tau\sim0.02-0.03$ in this wavelength region.

Fig.~\ref{f:6_68um} shows that the laboratory pure H$_2$O ice spectrum
generally does not explain all absorption in the 5-7 \mum\ region.  As
in \citet{boo08, boo11}, the residual 6.0 \mum\ absorption is fitted
with the empirical C1 and C2 components, and the 6.85 \mum\ absorption
with the components C3 and C4. The signal-to-noise ratios are low, and
no evidence is found for large C2/C1 or C4/C3 peak depth ratios,
which, towards YSOs, have been associated with heavily processed ices
\citep{boo08}.  Also, no evidence is found for the overarching C5
component, also possibly associated with energetic processing, at a
peak optical depth of $\leq 0.03$.


\begin{figure}[b]
\includegraphics[angle=90, scale=0.62]{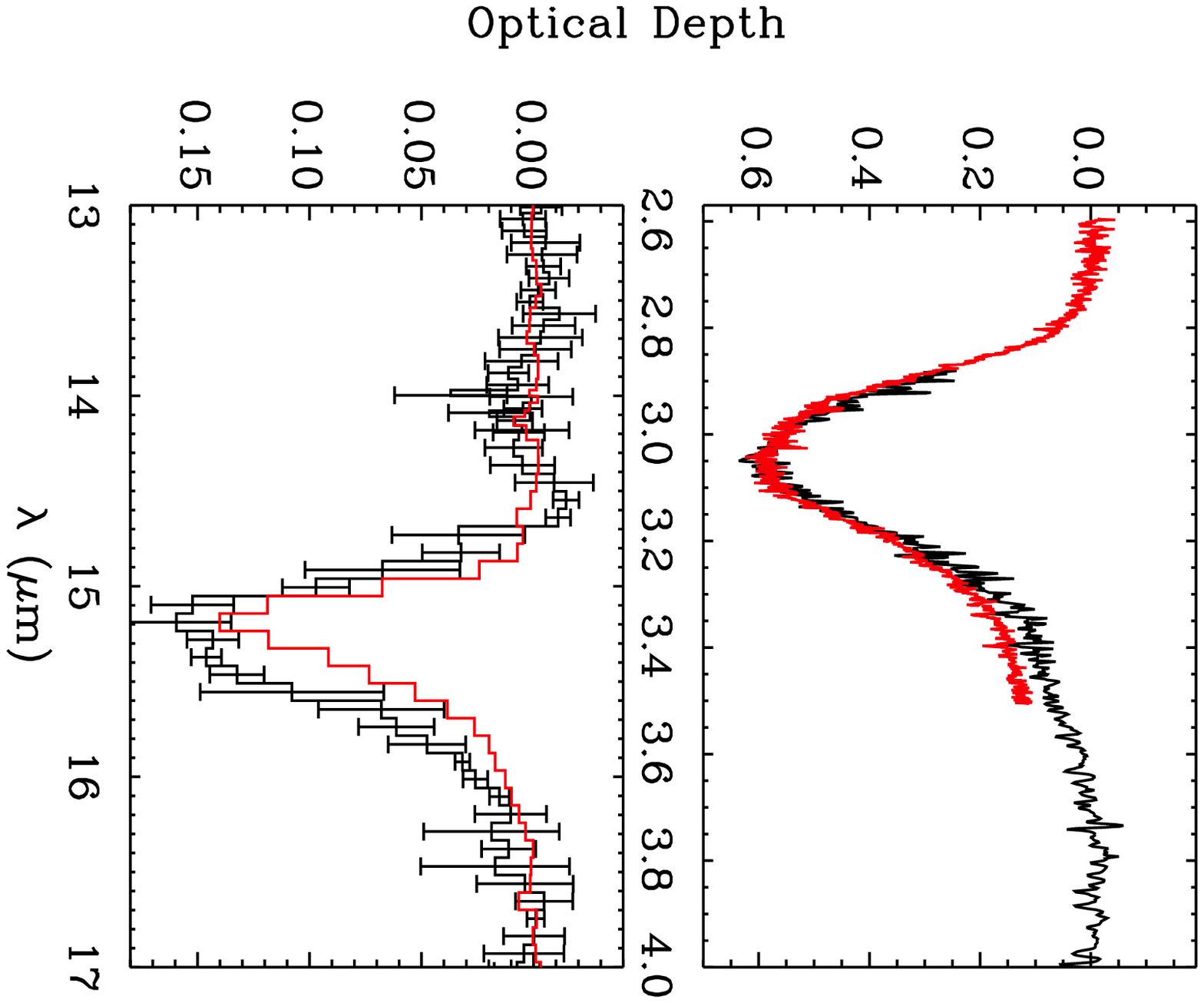}
\caption{3.0 \mum\ H$_2$O (top panel) and 15 \mum\ CO$_2$ (bottom)
  bands for the background stars 2MASS J15452747-3425184 (Lupus I; in
  black) and Elias 16 (TMC; in red). The spectra are scaled to the
  solid H$_2$O column density. The pronounced wing in the CO$_2$
  bending mode toward 2MASS J15452747-3425184 may indicate a larger
  fraction of H$_2$O-rich ices (not further analyzed in this
  work).}~\label{f:co2}
\end{figure}

\begin{deluxetable*}{lcccccc}
\tabletypesize{\footnotesize}
\setlength{\tabcolsep}{0.05in} 
\tablecolumns{7}
\tablewidth{0pc}
\tablecaption{Optical Depths 5-8~\mum\ Features~\label{t:tau}}
\tablehead{
\colhead{Source}& \multicolumn{4}{c}{$\tau_{\rm int}$ [cm$^{-1}$]}   &
                  \colhead{$\tau_{\rm 6.0}$\tablenotemark{e}}      &\colhead{$\tau_{\rm 6.85}$\tablenotemark{f}}\\
\colhead{      }&\colhead{5.2-6.4~\mum\tablenotemark{a}}&\colhead{5.2-6.4~\mum\tablenotemark{b}}&
                  \colhead{6.4-7.2~\mum\tablenotemark{c}}&\colhead{6.4-7.2~\mum\tablenotemark{d}}&
                  \colhead{ }      &\colhead{ }\\
\colhead{2MASS~J}&\colhead{               } & \colhead{ minus H$_2$O  } &  
                  \colhead{               } & \colhead{ minus H$_2$O  } & 
                  \colhead{               } & \colhead{               } }
\startdata
$16014254-4153064$  &  16.55   (1.17) &   5.04   (1.17) &  10.61   (1.90) &   5.87   (1.90) &  0.102   (0.023) &  0.083   (0.045) \\
$16004739-4203573$  &  10.72   (1.29) &   3.70   (1.29) &   3.60   (1.70) &   0.69   (1.70) &  0.081   (0.027) &  0.041   (0.038) \\
$16010642-4202023$  &  12.48   (1.40) &   4.21   (1.40) &   7.68   (2.09) &   4.28   (2.09) &  0.092   (0.031) &  0.061   (0.050) \\
$15423699-3407362$  &  10.80   (1.16) &   1.53   (1.16) &   7.86   (1.41) &   4.04   (1.41) &  0.073   (0.023) &  0.056   (0.036) \\
$16012635-4150422$  &  12.16   (1.37) &   4.23   (1.37) &   9.23   (1.98) &   5.96   (1.98) &  0.090   (0.027) &  0.068   (0.049) \\
$15452747-3425184$  &   9.76   (0.47) &   2.71   (0.47) &   4.32   (0.24) &   1.42   (0.24) &  0.070   (0.006) &  0.038   (0.005) \\
$16012825-4153521$  &   6.71   (1.25) &   0.79   (1.25) &   5.16   (1.41) &   2.72   (1.41) &  0.051   (0.023) &  0.047   (0.035) \\
$16021102-4158468$  &   3.07   (1.05) &   1.84   (1.05) &  -0.07   (1.39) &  -0.58   (1.39) &  0.026   (0.021) &  0.003   (0.035) \\
$16014426-4159364$  &   3.20   (1.10) &   1.41   (1.10) &   3.01   (1.49) &   2.27   (1.49) &  0.022   (0.024) &  0.036   (0.039) \\
$16022128-4158478$  &   5.87   (1.27) &   2.85   (1.27) &   3.13   (1.30) &   1.89   (1.30) &  0.045   (0.019) &  0.026   (0.031) \\
$16015887-4141159$  &  -1.12   (0.62) &  -2.91   (0.62) &   0.91   (0.75) &   0.18   (0.75) & -0.002   (0.008) &  0.015   (0.017) \\
$16023370-4139027$  &   0.49   (1.31) &  -1.41   (1.31) &   4.18   (1.43) &   3.39   (1.43) &  0.012   (0.023) &  0.043   (0.039) \\
$16023789-4138392$  &   0.64   (1.06) &  -0.81   (1.06) &   1.89   (1.74) &   1.29   (1.74) & -0.001   (0.022) &  0.017   (0.043) \\
$16004226-4146411$  &   2.08   (0.98) &   1.18   (0.98) &   0.74   (1.31) &   0.38   (1.31) &  0.015   (0.020) &  0.017   (0.033) \\
$16000067-4204101$  &   2.14   (1.04) &   0.91   (1.04) &   1.32   (1.50) &   0.81   (1.50) &  0.014   (0.023) &  0.022   (0.037) \\
$16011478-4210272$  &  -0.46   (0.66) &  -0.79   (0.66) &   0.48   (0.94) &   0.34   (0.94) &  0.001   (0.014) &  0.009   (0.024) \\
$16005559-4159592$  &   1.15   (1.17) &   0.48   (1.17) &   2.50   (1.63) &   2.22   (1.63) &  0.010   (0.024) &  0.026   (0.041) \\
$16000874-4207089$  &   0.34   (0.87) &   0.00   (0.87) &   0.98   (1.15) &   0.84   (1.15) &  0.013   (0.018) &  0.024   (0.029) \\
$16022921-4146032$  &  -0.20   (1.08) &  -0.65   (1.08) &  -0.43   (1.23) &  -0.61   (1.23) &  0.012   (0.021) &  0.006   (0.030) \\
$15382645-3436248$  &   0.71   (1.14) &   0.37   (1.14) &   1.20   (1.25) &   1.06   (1.25) &  0.012   (0.021) &  0.018   (0.030) \\
$15595783-4152396$  &   0.95   (0.75) &  -0.39   (0.75) &   0.76   (1.08) &   0.20   (1.08) &  0.015   (0.016) &  0.016   (0.029) \\
\enddata
\tablecomments{The sources are sorted in order of decreasing $A_{\rm
    K}$ values (Table~\ref{t:fits}). Uncertainties in parentheses based
  on statistical errors in the spectra only, unless noted otherwise
  below.}
\tablenotetext{a}{ integrated optical depth between 5.2-6.4~\mum\ in wavenumber units}
\tablenotetext{b}{ integrated optical depth between 5.2-6.4~\mum\ in wavenumber units, after subtraction of a laboratory spectrum of pure H$_2$O ice}
\tablenotetext{c}{ integrated optical depth between 6.4-7.2~\mum\ in wavenumber units}
\tablenotetext{d}{ integrated optical depth between 6.4-7.2~\mum\ in wavenumber units, after subtraction of a laboratory spectrum of pure H$_2$O ice}
\tablenotetext{e}{ peak optical depth at 6.0~\mum}
\tablenotetext{f}{ peak optical depth at 6.85~\mum}
\end{deluxetable*}

\subsubsection{15 \mum\ CO$_2$ Band}

The CO$_2$ bending mode at 15 \mum\ was detected at $>3\sigma$
significance in one line of sight (Table~\ref{t:colden};
Fig.~\ref{f:co2}).  Toward 2MASS J15452747-3425184 (Lupus I), the
CO$_2$/H$_2$O column density ratio is $44.9\pm 5.5\%$. Taking into
account the large error bars, only one other line of sight has a
significantly different CO$_2$/H$_2$O ratio: 2MASS J15450300-3413097 at
$18.4\pm 8.4$\%.

\subsubsection{4.7 \mum\ CO Band}

\begin{figure}[h]
\includegraphics[angle=90, scale=0.62]{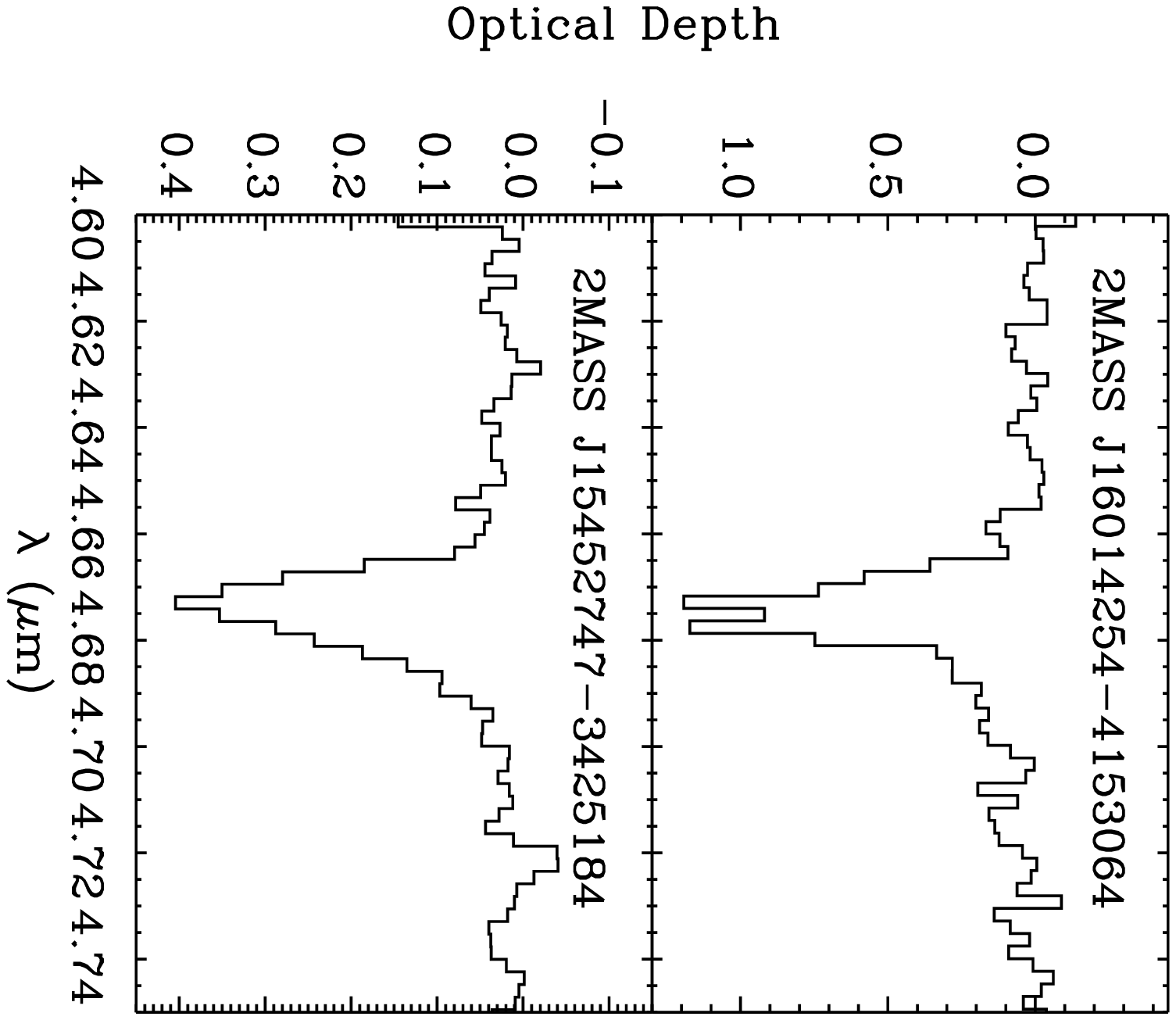}
\caption{The two Lupus lines of sight in which the solid CO band was
  detected.}~\label{f:co}
\end{figure}

The CO stretch mode at 4.7 \mum\ was detected at $>3\sigma$
significance in two lines of sight (out of five observed sight-lines),
one toward Lupus I and one toward Lupus IV (Table~\ref{t:colden}).
The detections are shown in Fig.~\ref{f:co}. The abundance relative to
H$_2$O is high, and significantly different between the two
detections: 42\% toward the Lupus IV source, and 26\% toward Lupus I.

\subsubsection{H$_2$CO and CH$_3$OH}~\label{sec:ch3oh}

Solid H$_2$CO and CH$_3$OH are not detected toward the Lupus
background stars. For CH$_3$OH, the 3.53 \mum\ C$-$H and the 9.7 O$-$H
stretch modes were used to determine upper limits to the column
density. Despite the overlap with the 9.7 \mum\ band of silicates, the
O$-$H stretch mode sometime gives the tightest constraint, because the
3.53 \mum\ region is strongly contaminated by narrow photospheric
absorption lines. The lowest upper limit of
$N$(CH$_3$OH)/$N$(H$_2$O)$<2.8$\% (3$\sigma$) is found for
2MASS J15452747-3425184. Other lines of sight have 3$\sigma$ upper
limits of 6-8\%, but larger if $N$(H$_2$O)$<4.5~10^{18}$ \sqcm.  For
H$_2$CO, the tightest upper limits are set by the strong C$=$O stretch
mode at 5.81 \mum: 4-6\% for lines of sight with the highest H$_2$O
column densities.

\begin{figure*}[t]
\includegraphics[angle=90, scale=0.75]{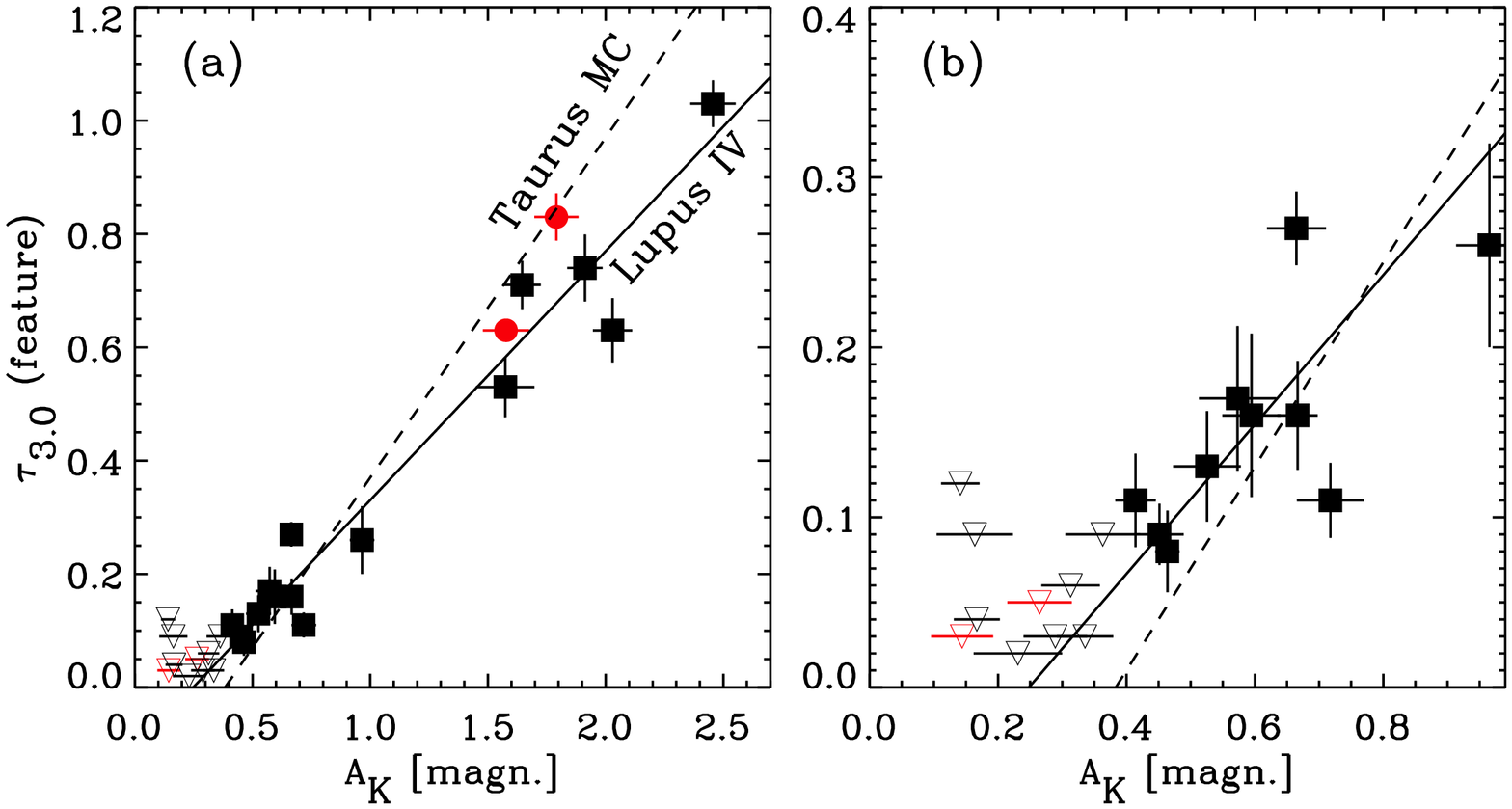}
\caption{{\bf Panel (a):} correlation plot of $A_{\rm K}$ with
  $\tau_{3.0}$. Background stars tracing the Lupus I cloud are
  indicated in red bullets and those tracing Lupus IV in black
  squares. Error bars are of 1$\sigma$ significance. Open triangles
  indicate 3$\sigma$ upper limits.  The solid line is a least-square
  fit to all Lupus IV detections. The dashed line represents the
  Taurus Molecular Cloud from \citet{whi01}. {\bf Panel (b):} same as
  panel (a), but highlighting the low extinction
  sight-lines.}~\label{f:corr11}
\end{figure*}

\subsubsection{HCOOH, CH$_4$, NH$_3$}~\label{sec:hcooh}

The spectra of the Lupus background stars were searched for signatures
of solid HCOOH, CH$_4$, and NH$_3$.  The absorption features were not
found, however, and for the sight-lines with the highest H$_2$O column
densities, the abundance upper limits are comparable or similar to the
limits for the isolated core background stars \citep{boo11}.  The 7.25
\mum\ C-H deformation mode of HCOOH, in combination with the 5.8
\mum\ C=O stretch mode, yields upper limits comparable to the typical
detections toward YSOs of 2-5\% relative to H$_2$O \citep{boo08}. The
7.68 \mum\ bending mode of CH$_4$ yields upper limits that are
comparable to the detections of 4\% toward YSOs \citep{obe08}.
Finally, for the NH$_3$ abundance, the 8.9 \mum\ umbrella mode yields
3$\sigma$ upper limits that are well above 20\% relative to H$_2$O
(Table 6), except for two lines of sight (2MASS J16012825−4153521 and
2MASS J16004739−4203573) which have 10\% upper limits.  These numbers
are not significant compared to the detections of 2-15\% toward YSOs
\citep{bot10}.




\subsection{Correlation Plots}~\label{sec:60}

The relationships between the total continuum extinction ($A_{\rm K}$)
and the strength of the H$_2$O ice ($\tau_{3.0}$) and silicates
($\tau_{9.7}$) features were studied in clouds and cores (e.g.,
\citealt{whi01, chi07,chi11,boo11}). Here they are derived for the
first time for the Lupus clouds.

\subsubsection{$\tau_{3.0}$ versus $A_{\rm K}$ }~\label{sec:tau30_ak}

The peak optical depth of the 3.0~\mum\ H$_2$O ice band correlates
well with $A_{\rm K}$ (Fig.~\ref{f:corr11}).  The Lupus I data points
(red bullets) are in line with those of Lupus IV. Still, these are
quite different environments (\S\ref{sec:intro}), and a linear fit is
only made to the Lupus IV detections, taking into account error bars
in both directions:

\begin{equation}
\tau_{3.0}=(-0.11\pm 0.03)+(0.44\pm 0.03)\times A_{\rm  K}~\label{eq:tau30ak}
\end{equation}

\noindent This relation implies a $\tau_{3.0}=0$ cut-off value of
$A_{\rm K}=0.25\pm 0.07$, which is the ``ice formation threshold''
further discussed in \S\ref{sec:thresh}. The lowest extinction at
which an ice band has been detected (at 3$\sigma$ level) is $A_{\rm
  K}$=0.41$\pm$0.03 mag. Most data points fall within 3$\sigma$ of
the linear fit. Two exceptions near $A_{\rm K}\sim$0.65 mag, and one
near 2.0 mag show that a linear relation does not apply to all Lupus
IV sight-lines.

\begin{figure}[h]
\includegraphics[angle=90, scale=0.58]{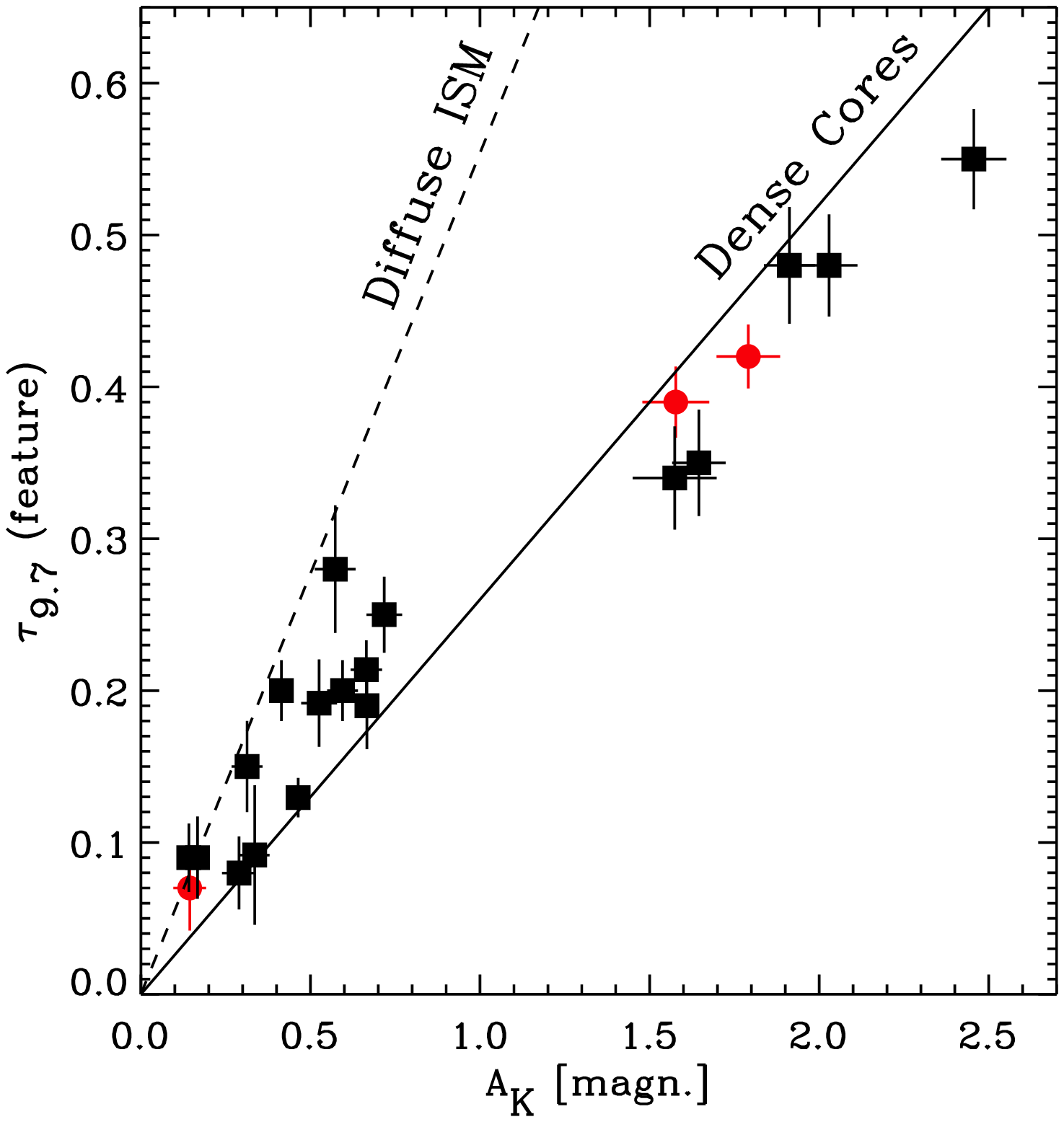}
\caption{Correlation plot of $A_{\rm K}$ with $\tau_{9.7}$. Background
  stars tracing the Lupus I cloud are indicated in red bullets and
  those tracing Lupus IV in black squares. Error bars are of 1$\sigma$
  significance. Sources with poor fits of the photospheric 8.0
  \mum\ SiO band (Table~\ref{t:fits}) are excluded from this plot. The
  dashed line is the diffuse medium relation taken from \citet{whi03},
  while the solid line is the relation for dense cores re-derived from
  \citet{boo11}. }~\label{f:corr1}
\end{figure}

\subsubsection{$\tau_{9.7}$ versus $A_{\rm K}$ }~\label{sec:tau97_ak}

The relation of $\tau_{9.7}$ with $A_{\rm K}$ was studied both in
diffuse \citep{whi03} and dense clouds \citep{chi07, boo11}.  For the
Lupus clouds, the data points are plotted in Fig.~\ref{f:corr1}.
Rather than fitting the data, the Lupus data are compared to the
distinctly different relations for the diffuse medium (\citealt{whi03};
dashed line in Fig.~\ref{f:corr1}):

\begin{equation}
\tau_{9.7}=0.554\times A_{\rm K}~\label{eq:tau97akdism}
\end{equation}

\noindent and the dense medium (solid line in Fig.~\ref{f:corr1}):

\begin{equation}
\tau_{9.7}=(0.26\pm 0.01) \times A_{\rm K}~\label{eq:tau97ak}
\end{equation}

\noindent The dense medium relation is re-derived from the isolated
dense core data points in \citet{boo11}, by forcing it through the
origin of the plot, and taking into account uncertainties in both
directions. To limit contamination by diffuse foreground dust, only
data points with $A_{\rm K}>1.4$ mag were included, and the L328 core
was excluded.

Fig.~\ref{f:corr1} shows that the Lupus lines of sight with $A_{\rm
  K}>$1.0 mag follow a nearly linear relation, though systematically
below the dense core fit.  At lower extinctions all sources scatter
rather evenly between the dense and diffuse medium relations. It is
worthwhile to note that none of the latter sources lie above the
diffuse or below the dense medium relations.




\section{Discussion}~\label{sec:dis}

\subsection{Ice Formation Threshold}~\label{sec:thresh}

The cut-off value of the relation between $A_{\rm K}$ and $\tau
_{3.0}$ fitted in Eq.~\ref{eq:tau30ak} and plotted in
Fig.~\ref{f:corr11} is referred to as the ``ice formation threshold''.
The Lupus IV cloud threshold of $A_{\rm K}=0.25\pm 0.07$ corresponds
to $A_{\rm V}=2.1\pm 0.6$.  Here, a conversion factor of $A_{\rm
  V}/A_{\rm K}$=8.4 is assumed, which is taken from the mean
extinction curves of \citet{car89} with $R_{\rm V}=3.5$, typical for
the lowest extinction lines of sight (\S\ref{sec:cont}).  The
threshold may be as low as 1.6 mag when taking into account the
contribution of diffuse foreground dust. \citet{knu98} derive a
contribution of $A_{\rm V}=0.12$ mag for distances up to 100 pc, but
for the Lupus IV cloud a foreground component at 50 pc with $A_{\rm
  V}\sim 0.5$ might be present.  Regardless of the foreground
extinction correction, the Lupus ice formation threshold is low
compared to that observed in other clouds and cores.  The difference
is at the 2$\sigma$ level compared to TMC ($A_{\rm V}=3.2\pm 0.1$
mag; \citealt{whi01}), but much larger compared to the Oph cloud
(10-15 mag; \citealt{tan90}).

The existence of the ice formation threshold and the differences
between clouds are a consequence of desorption (e.g., \citealt{wil92,
  pap05, cup07, hol09, caz10}). \citet{hol09} modeled the ice mantle
growth as a function of $A_{\rm V}$, taking into account photo, cosmic
ray, and thermal desorption, grain surface chemistry, an external
radiation field $G_0$, and time dependent gas phase chemistry.  At
high UV fields, the thermal (dust temperature) and photodesorption
rates are high, the residence time of H and O atoms on the grains is
short and the little H$_2$O that is formed will desorb rapidly.
Inside the cloud, dust attenuates the UV field and the beginnings of
an ice mantle are formed.  In these models, the extinction threshold
$A_{\rm Vf}$ is defined as the extinction at which the H$_2$O ice
abundance starts to increase rapidly, i.e., once a monolayer of ice is
formed, and desorption can no longer keep up with H$_2$O formation:

\begin{equation}
A_{\rm Vf}\propto {\rm ln}(G_0Y/n)\label{eq:avf}
\end{equation}

\noindent with $Y$ the photodesorption yield determined in laboratory
experiments and $n$ the gas density. For $Y=3~10^{-3}$, $G_0$=1, and
$n=10^3$ \cubcm, \citet{hol09} calculate $A_{\rm Vf}$=2 mag. To
compare this with the observations, this must be doubled because
background stars trace both the front and back of the cloud.  The
Lupus threshold is 1-2 mag lower than this calculation, but still
within the model uncertainties considering that $Y$ is not known with
better than 60\% accuracy \citep{obe09}. Also, the Lupus clouds may
have a lower radiation field (there are no massive stars closeby) or a
higher density.  On the other hand, the much higher threshold for the
Oph cloud is likely caused by the high radiation field from nearby hot
stars.  Shocks and radiation fields generated by the high SFR, or the
high mean stellar mass in the Oph cloud (\S\ref{sec:intro}) may play a
role as well, but the models of \citet{hol09} do not take this into
account. Indeed, the SFR and mean stellar mass of YSOs are low within
Lupus \citep{mer08,eva09}.

Three Lupus IV lines of sight deviate more than 3$\sigma$ from the
linear fit to the $A_{\rm K}$ versus $\tau _{3.0}$ relation
(Fig.~\ref{f:corr11}; \S\ref{sec:tau30_ak}). The TMC relation shows no
significant outliers \citep{whi01, chi11}, which is reflected in a low
uncertainty in the ice formation threshold.  A much larger scatter is
observed toward the sample of isolated cores of \citet{boo11}, which
likely reflects different ice formation thresholds toward different
cores or different contributions by diffuse ISM foreground dust
absorption.  For Lupus, the scatter may be attributed to the spread
out nature of the cloud complex. External radiation may penetrate
deeply in between the relatively small individual Lupus clouds and
clumps, in contrast to the TMC1 cloud, which is larger and more
homogeneous in the extinction maps of \citet{cam99} (the TMC and Lupus
cloud distances are both $\sim$150 pc).

\subsection{Slope of $\tau_{3.0}$ versus $A_{\rm K}$ Relation}~\label{sec:slope}

The slope in the relation between $A_{\rm K}$ and $\tau _{3.0}$ is a
measure of the H$_2$O ice abundance, which can be considered an
average of the individual abundances listed in Table~\ref{t:colden}.
Deep in the cloud ($A_{\rm V}=4$), a linear relation is expected for a
constant abundance, as most oxygen is included in H$_2$O
\citep{hol09}. The conversion factor between the slope in
Eq.~\ref{eq:tau30ak} and $x{\rm (H_2O)}$ is 5.23$\times 10^{-5}$.
This follows from Eq.~\ref{eq:nh2} and from $N({\rm
  H_2O})$=$\tau_{3.0} \times 322/2.0\times 10^{-16}$, where the
numerator is the $FWHM$ width of the 3.0 \mum\ band in cm$^{-1}$ and
the denominator the integrated band strength of H$_2$O ice in units of
cm/molecule.  This yields $x{\rm (H_2O)}=2.3\pm 0.1\times 10^{-5}$ for
Lupus. The error bar reflects the point to point scatter. The absolute
uncertainty is much larger, e.g., $\sim 30\%$ due to the effect of
$R_{\rm V}$ uncertainties on Eq.~\ref{eq:nh2}. For TMC, the slope is
steeper, which is illustrated in Fig.~\ref{f:corr11}, where the
relation of \citet{whi01} has been converted to an $A_{\rm K}$ scale
assuming $A_{\rm V}/A_{\rm K}$=8.4:

\begin{equation}
\tau_{3.0}=(-0.23\pm 0.01)+(0.60\pm 0.02)\times A_{\rm  K}~\label{eq:tau30aktmc}
\end{equation}

\noindent 
This translates to $x{\rm (H_2O)}=3.1\pm 0.1\times 10^{-5}$, which is
$\sim$35\% larger compared to Lupus. An entirely different explanation
for the slope difference may be the $A_{\rm V}/A_{\rm K}$ conversion
factor.  $A_{\rm V}/A_{\rm K}\sim 7.0$ would reduce the TMC slope to
the one for Lupus. However, \citet{whi01} use the relation $A_{\rm
  V}=5.3E_{\rm J-K}$ to convert infrared extinction to $A_{\rm V}$.
Using the mean extinction curve of \citet{car89}, this corresponds to
$A_{\rm V}/A_{\rm K}=9.3$, which increases the slope difference.  A
direct determination of $A_{\rm V}/A_{\rm K}$ at high extinction would
be needed to investigate these discrepancies. Alternatively, it is
recommended that inter-cloud comparisons of the $\tau_{3.0}$ formation
threshold and growth are done on the same ($A_{\rm K}$) scale.

\subsection{Ice Abundances and Composition}~\label{sec:complex}

The ice abundances in the Lupus clouds (Table~\ref{t:colden} and
\S\ref{sec:slope}) are similar to other quiescent lines of
sight. Applying Eq.~\ref{eq:nh2} to determine $x{\rm (H_2O)}$ for the
sample of isolated dense cores of \citet{boo11} yields values of
1.5-3.4$\times 10^{-5}$ (Table~\ref{t:appbg} in
Appendix~\ref{sec:app}), i.e., the Lupus abundances are within this
narrow range. For the sample of YSOs of \citet{boo08}, $x{\rm (H_2O)}$
may be determined from Eqs.~\ref{eq:nh2} and~\ref{eq:tau97ak},
yielding values of 0.6-7$\times 10^{-5}$ (Table~\ref{t:appyso} in
Appendix~\ref{sec:app}). The lowest abundances are seen toward YSOs
with the warmest envelopes \citep{tak00}, while the highest abundances
tend to be associated with more embedded YSOs. A high abundance of
8.5$\times 10^{-5}$ was also found in the inner regions of a Class 0
YSO \citep{pon04}.  Thus, whereas the ice abundance is remarkably
constant in quiescent dense clouds, it is apparently not saturated, as
it increases with a factor of 3-4 in dense YSO envelopes.

Of the upper limits determined for CH$_3$OH, NH$_3$, CH$_4$, and HCOOH
(\S\S\ref{sec:ch3oh} and~\ref{sec:hcooh}), the ones for CH$_3$OH are
most interesting. While they are comparable to the upper limits
determined in many other quiescent lines of sight in cores and clouds
\citep{chi95, boo11}, the lowest upper limits ($<2.8$\%) are
significantly below the CH$_3$OH abundances in several isolated cores
($\sim 10\%$; \citealt{boo11}).  In the scenario that CH$_3$OH is
formed by reactions of atomic H with frozen CO \citep{cup09}, this
indicates that the gas phase H/CO abundance ratio is rather low in the
Lupus clouds. This may be explained by a high density (promoting H$_2$
formation; \citealt{hol71}), or a low CO ice abundance.  The latter
may be a consequence of young age as CO is still being accreted. A
high dust temperature ($>15$ K) would slow down the accretion as well.
For only two Lupus lines of sight has solid CO been detected, and
although their abundances are high (42 and 26\% w.r.t. H$_2$O;
Table~\ref{t:colden}), they are low compared to lines of sight with
large CH$_3$OH abundances ($\sim$100\% of CO in addition to $\sim$28\%
of CH$_3$OH; \citealt{pon04}).  Thus it appears that in the Lupus
clouds the CO mantles are still being formed and insufficient H is
available to form CH$_3$OH. At such early stage more H$_2$CO than
CH$_3$OH may be formed \citep{cup09}, but H$_2$CO was not detected
with upper limits that are above the CH$_3$OH upper limits
(\S\ref{sec:ch3oh}).

For only one embedded YSO in the Lupus clouds have ice abundances been
determined (Table~\ref{t:colden}). It is IRAS 15398-3359 (SSTc2d
J154301.3-340915; 2MASS J15430131-3409153) classified as a Class 0 YSO
based on its low bolometric temperature (\citealt{kri12}; although the
$K$-band to 24 \mum\ photometry slope is more consistent with a Flat
spectrum YSO; \citealt{mer08}). Its CH$_3$OH abundance
(10.3$\pm$0.8\%) is well above the 3$\sigma$ upper limit toward the
nearest background star in Lupus I ($<7.8\%$; 2MASS J15423699-3407362)
at a distance of 5.3 arcmin (48000 AU at the Lupus distance of
$\sim$150 pc; \citealt{com08}) at the edge of the same core
(Fig.~\ref{f:maps}). In the same grain surface chemistry model, this
reflects a larger gas phase H/CO ratio due to higher CO freeze out as
a consequence of lower dust temperatures and possibly longer time
scales within the protostellar envelope compared to the surrounding
medium.

\subsection{$\tau_{9.7}/A_{\rm K}$ Relation}~\label{sec:tauak}

\begin{figure}[t]
\includegraphics[angle=90, scale=0.50]{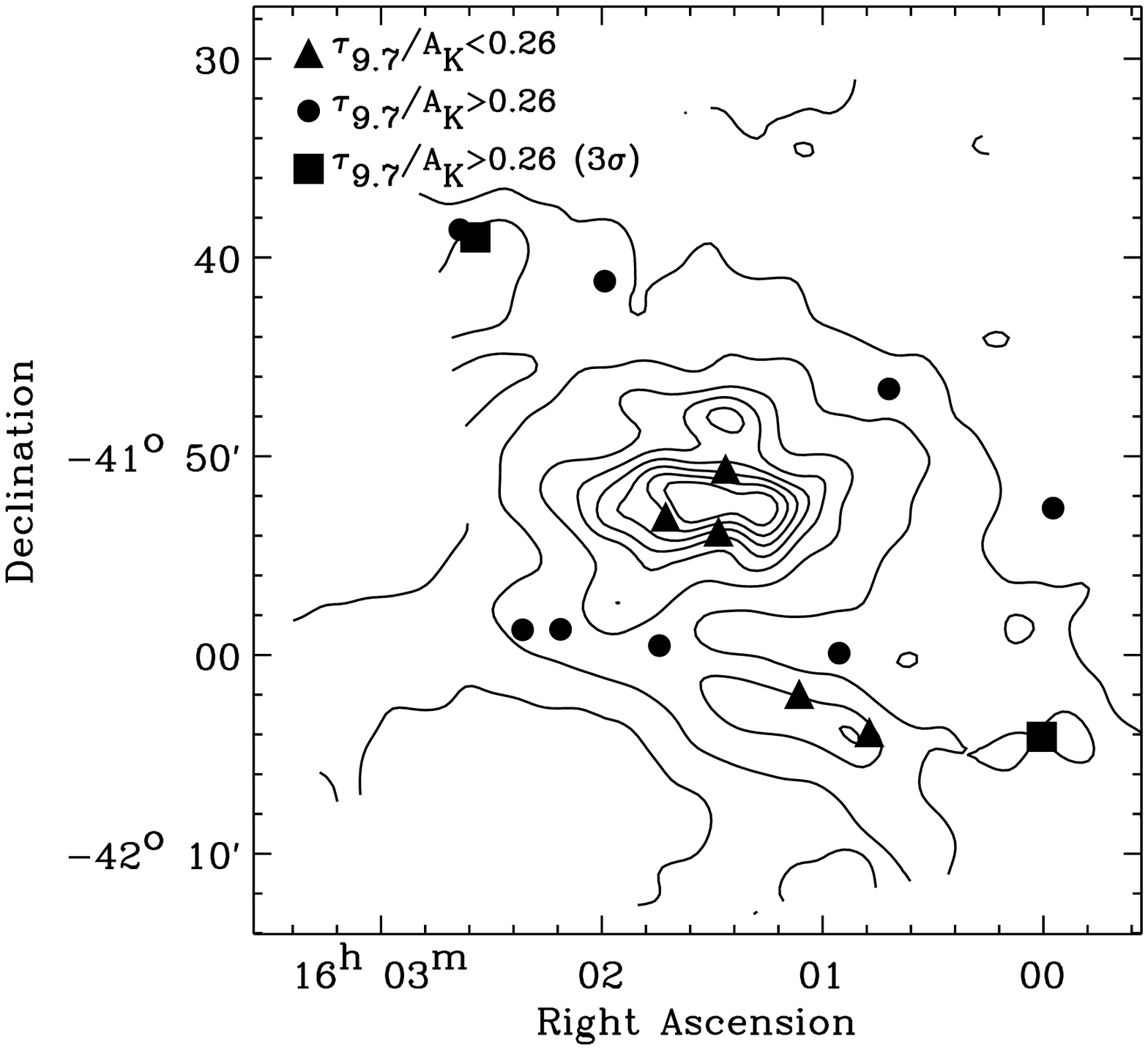}
\caption{Lupus IV lines of sight with $\tau_{9.7}/A_{\rm K}<0.26$
  (triangles) and $\tau_{9.7}/A_{\rm K}>0.26$ (bullets). Lines of
  sight that are more than 3$\sigma$ above $\tau_{9.7}/A_{\rm K}=0.26$
  are indicated with squares and have the most diffuse-medium type
  characteristics. The extinction contours \citep{eva07} represent
  $A_{\rm K}=0.25,0.5,1.0,1.5,...,3.5$ mag (assuming $A_{\rm V}/A_{\rm
    K}=7.6$), and have a spatial resolution of 2
  arcmin.}~\label{f:tau97akmap}
\end{figure}

Figure~\ref{f:corr1} shows that lines of sight through the Lupus
clouds with $A_{\rm K}> 1.0$ mag generally have the lowest
$\tau_{9.7}/A_{\rm K}$ ratios, i.e., they tend to lie below the dense
cores relation of \citet{boo11}.  This is further illustrated in
Fig.~\ref{f:tau97akmap}: Lupus IV sources with the lowest
$\tau_{9.7}/A_{\rm K}$ are concentrated in the highest extinction
regions. At lower extinctions ($A_{\rm K}< 1.0$ mag), the
$\tau_{9.7}/A_{\rm K}$ values scatter evenly between the diffuse
medium and dense core relations. Apparently, the transformation from
diffuse medium-type dust to dense medium-type dust is influenced by
the line of sight conditions. This is demonstrated by comparing the
two sight-lines 2MASS J16000067$-$4204101 and 16004226$-$4146411: at
comparable extinctions ($A_{\rm K}$=0.41 and 0.46;
Table~\ref{t:fits}), the first one follows the diffuse medium
$\tau_{9.7}/A_{\rm K}$ relation and the second one the dense medium
relation. Grain growth appears to play a role, as the latter has a
much larger $A_{7.4}/A_{\rm K}$ ratio (0.26 versus $<0.1$). Despite
its diffuse medium characteristics, 2MASS J16000067 has as much H$_2$O
ice as 2MASS J16004226 (Table~\ref{t:colden}).  In conclusion, the
process responsible for decreasing the $\tau_{9.7}/A_{\rm K}$ ratio in
the Lupus dense clouds is most likely related to grain growth, as was
also suggested by models \citep{bre11}. It is, however, not directly
related to ice mantle formation. Conversely, ice mantles may form on
grains before the process of grain coagulation has started.

\section{Conclusions and Future Work}~\label{sec:concl}

Photometry and spectroscopy at 1-25 \mum\ of background stars reddened
by the Lupus I and IV clouds is used to determine the properties of
the dust and ices before they are incorporated into circumstellar
envelopes and disks. The conclusions and directions for future work
are as follows:

\begin{enumerate}

\item H$_2$O ices form at extinctions of $A_{\rm K}=0.25\pm 0.07$ mag
  ($A_{\rm V}=2.1\pm 0.6$). This Lupus ice formation threshold is low
  compared to other clouds and cores, but still within 2$\sigma$ of
  the threshold in TMC. It is consistent with the absence of nearby
  hot stars that would photodesorb and sublimate the ices. To
  facilitate inter-cloud comparisons, independently from the applied
  optical extinction model, it is recommended to derive the threshold
  in $A_{\rm K}$ rather than $A_{\rm V}$.

\item The Lupus clouds are at an early chemical stage:

 \begin{itemize}

 \item The abundance of H$_2$O ice relative to $N_{\rm H}$ (2.3$\pm
   0.1 \times 10^{-5}$) is in the middle of the range found for other
   quiescent regions, but lower by a factor of 3-4 compared to dense
   envelopes of YSOs. The absolute uncertainty in the abundances,
   based on the uncertainty in $R_{\rm V}$ is estimated to be 30\%.

 \item While abundant solid CO is detected (26-42\% relative to
   H$_2$O), CO is not fully frozen out.

 \item The CH$_3$OH abundance is low ($<3-8$\% w.r.t. H$_2$O) compared
   to some isolated dense cores and dense YSO envelopes.  It indicates
   a low gas phase H/CO ratio, consistent with incomplete CO freeze
   out, possibly as a consequence of short time scales \citep{cup09}.

 \end{itemize}

\item The Lupus clouds have a low SFR and low stellar mass, thus
  limiting the effects of internal cloud heating and shocks on the ice
  abundances.  However, a larger diversity of clouds needs to be
  studied to determine the importance of star formation activity on
  the absolute and relative ice abundances in quiescent lines of
  sight. So far, high solid CH$_3$OH abundances were only found toward
  isolated dense cores, suggesting that the star formation environment
  (e.g., isolated versus clustered star formation) may play a role.

\item The spectra allow a separation of continuum extinction and ice
  and dust features, and continuum-only extinction curves are derived
  for different $A_{\rm K}$ values. Grain growth is evident in the
  Lupus clouds. More reddened lines of sight have larger mid-infrared
  ($>5$ \mum) continuum extinctions relative to $A_{\rm K}$.
  Typically, the Lupus background stars are best fitted with curves
  corresponding to $R_{\rm V}\sim 3.5$ ($A_{\rm K}=0.71$) and $R_{\rm
    V}\sim 5.0$ ($A_{\rm K}=1.47$).

\item The $\tau_{9.7}/A_{\rm K}$ ratio in Lupus is slightly less than
  that of isolated dense cores for lines of sight with $A_{\rm K}>1.0$
  mag, i.e., it is a factor of 2 lower compared to the diffuse
  medium. Below 1.0 mag values scatter between the dense and diffuse
  medium ratios. The absence of a gradual transition between diffuse
  and dense medium-type dust indicates that local conditions matter in
  the process that sets the $\tau_{9.7}/A_{\rm K}$ ratio. It is found
  that the reduction of $\tau_{9.7}/A_{\rm K}$ ratio in the Lupus
  dense clouds is most likely related to grain growth, which occurs in
  some sight-lines and not in others. It is, however, not directly
  related to ice mantle formation. Conversely, ice mantles may form on
  grains before the process of grain coagulation has started.  Future
  work needs to study the $\tau_{9.7}/A_{\rm K}$ ratio at $A_{\rm
    K}<1.0$ mag in more detail in both the dense and diffuse ISM to
  address the conditions that set the transition between the two
  environments.

\item All aspects of this work will benefit from improved stellar
  models.  Current models often do not simultaneously fit the
  strengths of the 2.4 \mum\ CO overtone band, the CO fundamental near
  5.3 \mum, and the SiO band near 8.0 \mum. In addition, the search for
  weak ice bands, such as that of CH$_3$OH at 3.53 \mum\ band, is
  limited by the presence of narrow photospheric lines.  Correction for
  photospheric lines will become the limiting factor in high S/N
  spectra at this and longer wavelengths and higher spectral
  resolution with new facilities (SOFIA, JWST, TMT).

\end{enumerate}


\acknowledgments 

We thank the anonymous referee for detailed comments that improved the
presentation of the results.  This work is based on observations made
with the Spitzer Space Telescope, which is operated by the Jet
Propulsion Laboratory (JPL), California Institute of Technology
(Caltech) under a contract with the National Aeronautics and Space
Administration (NASA). Support for this work was provided by NASA
through awards issued by JPL/Caltech to JEC and CK. This publication
makes use of data products from the Wide-field Infrared Survey
Explorer, which is a joint project of the University of California,
Los Angeles, and JPL/Caltech, funded by NASA.  This publication makes
use of data products from the Two Micron All Sky Survey, which is a
joint project of the University of Massachusetts and the Infrared
Processing and Analysis Center/Caltech, funded by NASA and the
National Science Foundation.

\appendix
\section{H$_2$O Ice Abundances YSOs and Isolated Dense Cores}~\label{sec:app}

Previous ice surveys generally list H$_2$O ice column densities and
abundances relative to H$_2$O ice, but not the abundances of H$_2$O
with respect to the hydrogen column density $N_{\rm H}=N{\rm
  (H)}+N{\rm (H_2)}$. Here, analogous to the Lupus background stars
(\S\ref{sec:h2o}; Table~\ref{t:colden}), H$_2$O abundances

\begin{equation}
x{\rm (H_2O)}=N{\rm (H_2O)}/N_{\rm H}~\label{eq:abun}
\end{equation}

are derived for the background stars of isolated dense cores of
\citet{boo11} and for the YSOs of \citet{boo08}. For the background
stars, $N_{\rm H}$ was calculated using Eq.~\ref{eq:nh2} of this work
and the $A_{\rm K}$ values of \citet{boo11}. A column of 4.0$\times
10^{21}$ \sqcm\ was subtracted from N$_{\rm H}$ before dividing $N{\rm
  (H_2O)}$ over N$_{\rm H}$, to correct for the ice formation
threshold, assuming that it has the same value as for Lupus
(\S\ref{sec:tau30_ak}). The results are listed in Table~\ref{t:appbg}.

For the YSOs of \citet{boo08}, Eq.~\ref{eq:nh2} was used as well to
determine $N_{\rm H}$, but $A_{\rm K}$ was not directly measured and
it was determined from the silicate band following
Eq.~\ref{eq:tau97ak}. The resulting $x{\rm (H_2O)}$ values are on one
hand underestimated because no correction is made for iceless grains
in the warm inner regions of the YSOs, and on the other hand
overestimated for several YSOs due to filling of the 9.7
\mum\ absorption band by emission. The results are listed in
Table~\ref{t:appyso}.

Neither in Table~\ref{t:appbg} nor \ref{t:appyso} is the uncertainty
in Eq.~\ref{eq:nh2} taken into account.  The effect of just the
uncertainty in $R_{\rm V}$ on the derived abundances is estimated to
be 30\% (\S\ref{sec:h2o}). A potentially larger uncertainty is that of
the $A_{\rm V}/N_{\rm H}$ ratio. Its accuracy is unknown as it was
determined in only one dense cloud (Oph; \citealt{boh78}).

\begin{deluxetable}{lllccc}
\tabletypesize{\footnotesize}
\setlength{\tabcolsep}{0.06in} 
\tablecolumns{6}
\tablewidth{0pc}
\tablecaption{Hydrogen Column Densities and H$_2$O Ice Abundances Isolated Dense Cores~\label{t:appbg}}
\tablehead{
\colhead{Source}& \colhead{Core\tablenotemark{a}}& \colhead{$N$(H$_2$O)}     & 
                  \colhead{$A_{\rm K}$}   &
                  \colhead{$N_{\rm H}$}   &
                  \colhead{$x$(H$_2$O)}  \\
\colhead{2MASS J}&\colhead{}& \colhead{$10^{18}$ } & 
                   \colhead{mag      } & 
                   \colhead{$10^{22}$ } & 
                   \colhead{$10^{-5}$ } \\
\colhead{       }&\colhead{}& \colhead{ \sqcm} & 
                   \colhead{      } & 
                   \colhead{ \sqcm} & 
                   \colhead{      } \\ }
\startdata
 $\rm           12014301-6508422        $ &   DC 297.7-2.8*  &      5.40 (1.45)\tablenotemark{b} &  5.17 ( 0.15 ) & 15.52 ( 0.48) &                3.48 (0.91)\tablenotemark{b} \\
 $\rm           18171366-0813188        $ &         L 429-C  &      3.40                  (0.38) &  3.58 ( 0.11 ) & 10.62 ( 0.33) &                3.20                  (0.35) \\
 $\rm           18172690-0438406        $ &          L 483*  &      4.31                  (0.48) &  4.60 ( 0.14 ) & 13.77 ( 0.42) &                3.13                  (0.35) \\
 $\rm           18170470-0814495        $ &         L 429-C  &      3.93                  (0.44) &  4.27 ( 0.13 ) & 12.75 ( 0.39) &                3.09                  (0.34) \\
 $\rm           18160600-0225539        $ &        CB 130-3  &      1.06                  (0.19) &  1.25 ( 0.04 ) &  3.44 ( 0.12) &                3.08                  (0.50) \\
 $\rm           18171181-0814012        $ &         L 429-C  &      3.81                  (0.42) &  4.31 ( 0.13 ) & 12.88 ( 0.40) &                2.96                  (0.33) \\
 $\rm           18170957-0814136        $ &         L 429-C  &      2.85                  (0.31) &  3.27 ( 0.10 ) &  9.69 ( 0.30) &                2.94                  (0.32) \\
 $\rm           04393886+2611266        $ &       Taurus MC  &      2.39                  (0.26) &  3.00 ( 0.09 ) &  8.85 ( 0.28) &                2.71                  (0.30) \\
 $\rm           12014598-6508586        $ &   DC 297.7-2.8*  &      2.44 (0.48)\tablenotemark{b} &  3.07 ( 0.09 ) &  9.07 ( 0.28) &                2.69 (0.51)\tablenotemark{b} \\
 $\rm           08093468-3605266        $ &       CG 30-31*  &      3.54 (0.79)\tablenotemark{b} &  4.52 ( 0.14 ) & 13.53 ( 0.42) &                2.61 (0.57)\tablenotemark{b} \\
 $\rm           18140712-0708413        $ &           L 438  &      1.40                  (0.15) &  1.89 ( 0.06 ) &  5.43 ( 0.17) &                2.59                  (0.27) \\
 $\rm           21240517+4959100        $ &          L 1014  &      2.19                  (0.24) &  3.10 ( 0.09 ) &  9.15 ( 0.29) &                2.39                  (0.26) \\
 $\rm           19214480+1121203        $ &         L 673-7  &      1.43                  (0.16) &  2.10 ( 0.06 ) &  6.06 ( 0.19) &                2.36                  (0.25) \\
 $\rm           22063773+5904520        $ &         L 1165*  &      1.13                  (0.12) &  1.74 ( 0.09 ) &  4.96 ( 0.27) &                2.28                  (0.25) \\
 $\rm           17160467-2057072        $ &          L 100*  &      2.13                  (0.23) &  3.24 ( 0.10 ) &  9.58 ( 0.30) &                2.23                  (0.24) \\
 $\rm           21240614+4958310        $ &          L 1014  &      0.98                  (0.11) &  1.60 ( 0.05 ) &  4.55 ( 0.15) &                2.16                  (0.23) \\
 $\rm           17160860-2058142        $ &          L 100*  &      1.53                  (0.17) &  2.45 ( 0.07 ) &  7.16 ( 0.23) &                2.14                  (0.23) \\
 $\rm           17155573-2055312        $ &          L 100*  &      1.63                  (0.18) &  2.65 ( 0.08 ) &  7.77 ( 0.24) &                2.10                  (0.23) \\
 $\rm           17112005-2727131        $ &           B 59*  &      3.79                  (0.42) &  6.00 ( 0.18 ) & 18.10 ( 0.55) &                2.09                  (0.23) \\
 $\rm           04215402+1530299        $ &     IRAM 04191*  &      1.84                  (0.20) &  3.04 ( 0.09 ) &  8.96 ( 0.28) &                2.05                  (0.22) \\
 $\rm           17111501-2726180        $ &           B 59*  &      2.35                  (0.26) &  3.91 ( 0.12 ) & 11.65 ( 0.36) &                2.02                  (0.22) \\
 $\rm           17111538-2727144        $ &           B 59*  &      2.01                  (0.22) &  3.53 ( 0.11 ) & 10.48 ( 0.33) &                1.92                  (0.21) \\
 $\rm           18165917-1801158        $ &           L 328  &      1.85                  (0.20) &  3.34 ( 0.10 ) &  9.88 ( 0.31) &                1.87                  (0.20) \\
 $\rm           18165296-1801287        $ &           L 328  &      1.62                  (0.22) &  3.00 ( 0.18 ) &  8.83 ( 0.55) &                1.83                  (0.26) \\
 $\rm           08093135-3604035        $ &       CG 30-31*  &      0.92 (0.42)\tablenotemark{b} &  1.89 ( 0.06 ) &  5.42 ( 0.17) &                1.71 (0.73)\tablenotemark{b} \\
 $\rm           15421699-5247439        $ &      DC 3272+18  &      1.43 (0.44)\tablenotemark{b} &  2.87 ( 0.09 ) &  8.43 ( 0.26) &                1.70 (0.50)\tablenotemark{b} \\
 $\rm           18170429-1802540        $ &           L 328  &      1.43                  (0.20) &  2.96 ( 0.09 ) &  8.72 ( 0.27) &                1.64                  (0.22) \\
 $\rm           18300061+0115201        $ &      Serpens MC  &      3.04                  (0.34) &  6.38 ( 0.19 ) & 19.24 ( 0.59) &                1.58                  (0.17) \\
 $\rm           18170426-1802408        $ &           L 328  &      1.09                  (0.12) &  2.50 ( 0.08 ) &  7.31 ( 0.23) &                1.50                  (0.16) \\
 $\rm           19201597+1135146        $ &         CB 188*  &      1.34                  (0.14) &  3.14 ( 0.09 ) &  9.29 ( 0.29) &                1.44                  (0.16) \\
 $\rm           19201622+1136292        $ &         CB 188*  &      1.01                  (0.11) &  2.50 ( 0.13 ) &  7.32 ( 0.39) &                1.38                  (0.16) \\
 $\rm           15421547-5248146        $ &      DC 3272+18  &   $<$1.18                         &  1.94 ( 0.06 ) &  5.59 ( 0.18) &             $<$2.11                         \\
 $\rm           08052135-3909304        $ &          BHR 16  &   $<$1.18                         &  1.47 ( 0.04 ) &  4.12 ( 0.14) &             $<$2.86                         \\
\enddata
\tablecomments{The rows are ordered in decreasing $x$(H$_2$O)}
\tablenotetext{a}{Cores with a '*' at the end of their name contain YSOs.}
\tablenotetext{b}{The H$_2$O column density and abundance are
  uncertain because no $L-$band spectra are available for this
  source.}
\end{deluxetable}

\begin{deluxetable}{llccc}
\tabletypesize{\footnotesize}
\setlength{\tabcolsep}{0.06in} 
\tablecolumns{5}
\tablewidth{0pc}
\tablecaption{Hydrogen Column Densities and H$_2$O Ice Abundances YSOs~\label{t:appyso}}
\tablehead{
\colhead{Source}& \colhead{$N$(H$_2$O)}     & 
                  \colhead{$A_{\rm K}$\tablenotemark{a}}   &
                  \colhead{$N_{\rm H}$}                    &
                  \colhead{$x$(H$_2$O)}  \\
\colhead{       }& \colhead{$10^{18}$ } & 
                   \colhead{mag } & 
                   \colhead{$10^{22}$ } & 
                   \colhead{$10^{-5}$ } \\
\colhead{       }& \colhead{ \sqcm} & 
                   \colhead{      } & 
                   \colhead{ \sqcm} & 
                   \colhead{      } \\ }
\startdata
 $\rm                 L1455 IRS3        $ &       0.75 (0.38)\tablenotemark{b} &  0.57 ( 0.07 ) &  1.36 ( 0.21) &     5.55    (2.25)\tablenotemark{b} \\
 $\rm            IRAS 03235+3004        $ &       12.1 (2.56)\tablenotemark{b} &  7.61 ( 0.77 ) & 22.95 ( 2.36) &     5.29    (1.21)\tablenotemark{b} \\
 $\rm           IRAS 04108+2803B        $ &       2.87                  (0.49) &  1.90 ( 0.19 ) &  5.43 ( 0.59) &     5.29                     (0.98) \\
 $\rm                     RNO 91        $ &       4.25                  (0.55) &  3.07 ( 0.31 ) &  9.02 ( 0.95) &     4.71                     (0.74) \\
 $\rm            IRAS 23238+7401        $ &       11.3 (2.53)\tablenotemark{b} &  8.84 ( 0.89 ) & 26.74 ( 2.74) &     4.26    (1.02)\tablenotemark{b} \\
 $\rm            IRAS 03271+3013        $ &       6.31 (1.87)\tablenotemark{b} &  4.97 ( 0.51 ) & 14.86 ( 1.56) &     4.24    (1.29)\tablenotemark{b} \\
 $\rm            IRAS 03245+3002        $ &       38.3 (6.82)\tablenotemark{b} & 31.21 ( 3.52 ) & 95.34 (10.80) &     4.02    (0.84)\tablenotemark{b} \\
 $\rm            IRAS 15398-3359        $ &       13.1 (4.16)\tablenotemark{b} & 12.77 ( 1.29 ) & 38.77 ( 3.94) &     3.40    (1.11)\tablenotemark{b} \\
 $\rm                       B1-c        $ &       27.9 (6.30)\tablenotemark{b} & 27.03 ( 2.80 ) & 82.53 ( 8.59) &     3.38    (0.83)\tablenotemark{b} \\
 $\rm     SSTc2dJ171122.2-272602        $ &       11.2 (3.03)\tablenotemark{b} & 11.50 ( 1.15 ) & 34.89 ( 3.54) &     3.21    (0.91)\tablenotemark{b} \\
 $\rm                 HH 100 IRS        $ &       2.45                  (0.34) &  2.64 ( 0.27 ) &  7.70 ( 0.83) &     3.18                     (0.52) \\
 $\rm                R CrA IRS 5        $ &       3.58                  (0.44) &  3.93 ( 0.39 ) & 11.67 ( 1.21) &     3.07                     (0.47) \\
 $\rm                    SVS 4-5        $ &       5.65                  (1.26) &  6.19 ( 0.63 ) & 18.61 ( 1.92) &     3.03                     (0.72) \\
 $\rm                      GL989        $ &       2.24                  (0.24) &  2.56 ( 0.26 ) &  7.44 ( 0.79) &     3.01                     (0.42) \\
 $\rm     2MASS J17112317-2724315       $ &       15.8 (3.57)\tablenotemark{b} & 17.30 ( 1.74 ) & 52.69 ( 5.33) &     3.00    (0.73)\tablenotemark{b} \\
 $\rm                  HH 46 IRS        $ &       7.71 (1.09)\tablenotemark{b} &  8.59 ( 0.86 ) & 25.95 ( 2.64) &     2.97    (0.50)\tablenotemark{b} \\
 $\rm                       B1-a        $ &       8.10 (2.40)\tablenotemark{b} &  9.03 ( 0.91 ) & 27.30 ( 2.79) &     2.96    (0.91)\tablenotemark{b} \\
 $\rm                    B5 IRS3        $ &       1.01                  (0.14) &  1.25 ( 0.13 ) &  3.43 ( 0.40) &     2.96                     (0.45) \\
 $\rm                  L1489 IRS        $ &       4.26                  (0.66) &  5.09 ( 0.51 ) & 15.22 ( 1.57) &     2.79                     (0.50) \\
 $\rm                 CrA IRS7 A        $ &       9.47 (2.14)\tablenotemark{b} & 11.63 ( 1.16 ) & 35.29 ( 3.57) &     2.68    (0.65)\tablenotemark{b} \\
 $\rm                    GL7009S        $ &       11.3                  (2.53) & 13.94 ( 1.44 ) & 42.37 ( 4.41) &     2.67                     (0.65) \\
 $\rm                 CrA IRS7 B        $ &       10.3 (2.23)\tablenotemark{b} & 13.22 ( 1.32 ) & 40.16 ( 4.06) &     2.57    (0.60)\tablenotemark{b} \\
 $\rm                  L1014 IRS        $ &       6.12 (1.10)\tablenotemark{b} &  7.88 ( 0.79 ) & 23.77 ( 2.43) &     2.57    (0.52)\tablenotemark{b} \\
 $\rm               NGC7538 IRS9        $ &       6.41                  (0.90) &  8.66 ( 0.87 ) & 26.17 ( 2.67) &     2.45                     (0.41) \\
 $\rm                 L1455 SMM1        $ &       14.4 (3.17)\tablenotemark{b} & 19.92 ( 2.20 ) & 60.72 ( 6.76) &     2.37    (0.58)\tablenotemark{b} \\
 $\rm            CRBR 2422.8-342        $ &       4.19                  (0.59) &  6.47 ( 0.65 ) & 19.46 ( 1.99) &     2.15                     (0.36) \\
 $\rm                     HH 300        $ &       2.59                  (0.36) &  4.20 ( 0.42 ) & 12.49 ( 1.29) &     2.08                     (0.34) \\
 $\rm                      EC 74        $ &       1.07                  (0.21) &  1.82 ( 0.18 ) &  5.18 ( 0.57) &     2.07                     (0.43) \\
 $\rm            IRAS 03254+3050        $ &       3.66                  (0.60) &  6.26 ( 0.63 ) & 18.80 ( 1.93) &     1.94                     (0.36) \\
 $\rm                   Elias 29        $ &       3.04                  (0.43) &  5.56 ( 0.56 ) & 16.65 ( 1.71) &     1.83                     (0.31) \\
 $\rm                       W33A        $ &       12.5                  (3.38) & 23.15 ( 2.55 ) & 70.61 ( 7.83) &     1.78                     (0.51) \\
 $\rm            IRAS 12553-7651        $ &       2.82 (0.63)\tablenotemark{b} &  5.39 ( 0.54 ) & 16.14 ( 1.65) &     1.75    (0.41)\tablenotemark{b} \\
 $\rm                 CrA IRAS32        $ &       4.89 (1.94)\tablenotemark{b} &  9.36 ( 1.07 ) & 28.32 ( 3.28) &     1.72    (0.70)\tablenotemark{b} \\
 $\rm                    B5 IRS1        $ &       2.26                  (0.36) &  4.75 ( 0.48 ) & 14.17 ( 1.46) &     1.59                     (0.29) \\
 $\rm                     GL2136        $ &       4.57                  (0.64) & 11.63 ( 1.16 ) & 35.30 ( 3.57) &     1.29                     (0.22) \\
 $\rm                   DG Tau B        $ &       2.29                  (0.45) &  7.87 ( 0.79 ) & 23.76 ( 2.42) &     0.96                     (0.21) \\
 $\rm                  S140 IRS1        $ &       1.95                  (0.27) &  7.43 ( 0.74 ) & 22.41 ( 2.28) &     0.87                     (0.14) \\
 $\rm                    W3 IRS5        $ &       5.65                  (0.80) & 24.15 ( 2.46 ) & 73.71 ( 7.55) &     0.76                     (0.13) \\
 $\rm            IRAS 03301+3111        $ &       0.40                  (0.06) &  1.90 ( 0.19 ) &  5.43 ( 0.58) &     0.73                     (0.12) \\
 $\rm                 MonR2 IRS3        $ &       1.59                  (0.22) &  8.66 ( 0.87 ) & 26.19 ( 2.66) &     0.60                     (0.10) \\
\enddata
\tablecomments{The rows are ordered in decreasing $x$(H$_2$O)}
\tablenotetext{a}{Determined from $\tau_{9.7}$ \citep{boo08}
  following Eq.~\ref{eq:tau97ak} from this work.}
\tablenotetext{b}{The H$_2$O column density and abundance are
  uncertain because no $L-$band spectra are available for this
  source.}
\end{deluxetable}

\end{document}